\documentclass[prd,showpacs,showkeys,floatfix,amsmath,amssymb,floatfix,english]{revtex4}

\usepackage{amsfonts}
\usepackage{subfig}
\usepackage{dcolumn}
\usepackage{natbib}
\usepackage{bm}
\usepackage{booktabs}
\usepackage{slashed}

\usepackage[T1]{fontenc}
\usepackage[utf8]{inputenc}
\usepackage{babel}
\usepackage{units}
\usepackage{amsmath}
\usepackage{esint}
\usepackage{multirow}
\usepackage{graphicx}
\usepackage[normalem]{ulem}
\usepackage[colorlinks, citecolor=blue,anchorcolor=red,menucolor=red, linkcolor=red,filecolor=red,runcolor=red,urlcolor=blue,frenchlinks=red]{hyperref}

\usepackage{ulem}
\usepackage{color}

\makeatletter
\newcommand{\figcaption}{\def\@captype{figure}\caption}
\newcommand{\tabcaption}{\def\@captype{table}\caption}

\newcommand{\Rmnum}[1]{\expandafter\@slowromancap\romannumeral #1@}

\def\hlinewd#1{%
  \noalign{\ifnum0=`}\fi\hrule \@height #1 \futurelet
   \reserved@a\@xhline}
\makeatother

\begin{document}

\title{Exotic $\bar{D}_s^{(*)}D^{(*)}$ molecular states and $sc\bar q\bar c$ tetraquark states with $J^P=0^+, 1^+, 2^+$}

\author{Qi-Nan Wang}
\author{Wei Chen}
\email{chenwei29@mail.sysu.edu.cn}
\affiliation{School of Physics, Sun Yat-Sen University, Guangzhou 510275, China}
\author{Hua-Xing Chen}
\email{hxchen@seu.edu.cn}
\affiliation{School of Physics, Southeast University, Nanjing 210094, China}

\begin{abstract}
We have calculated the mass spectra for the $\bar{D}_s^{(*)}D^{(*)}$ molecular states and $sc\bar q\bar c$ tetraquark states with $J^P=0^+, 1^+, 2^+$. 
The masses of the axial-vector $\bar{D}_sD^{*}$, $\bar{D}_s^{*}D$ molecular states and $\mathbf{1}_{[sc]} \oplus \mathbf{0}_{[\bar q \bar{c}]}$, $\mathbf{0}_{[sc]} \oplus \mathbf{1}_{[\bar q \bar{c}]}$ tetraquark states are predicted to be around 3.98 GeV, which are in good agreement with the mass of $Z_{cs}(3985)^-$ from BESIII \cite{besiii2020Zcs}.
In both the molecular and diquark-antidiquark pictures, our results suggest that there may exist two almost degenerate states, as the strange partners of the $X(3872)$ and $Z_c(3900)$. We propose to carefully examine the $Z_{cs}(3985)$ in future experiments to verify this. One may also search for more hidden-charm four-quark states with strangeness not only in the open-charm $\bar{D}_s^{(*)}D^{(*)}$ channels, but also in the hidden-charm channels $\eta_c K/K^\ast$, $J/\psi K/K^\ast$.  
\end{abstract}


\pacs{12.39.Mk, 12.38.Lg, 14.40.Ev, 14.40.Rt}
\keywords{molecular state, Exotic state, QCD sum rules}
\maketitle

\section{Introduction}
Very recently, the BESIII Collaboration announced a new structure near the $D_{s}^{-} D^{* 0}$ and $D_{s}^{*-} D^{0}$ thresholds in the $K^+$ recoil-mass spectra in $e^{+} e^{-} \rightarrow K^{+}\left(D_{s}^{-} D^{* 0}+D_{s}^{*-} D^{0}\right)$ \cite{besiii2020Zcs}. The pole mass and width of this $Z_{cs}(3985)^-$ resonance are measured as $\left(3982.5_{-2.6}^{+1.8} \pm 2.1\right) \mathrm{MeV} $ and $\left(12.8_{-4.4}^{+5.3} \pm 3.0\right) \mathrm{MeV}$, respectively. Decaying into $D_{s}^{-} D^{* 0}$ and $D_{s}^{*-} D^{0}$ in S-wave, the spin-parity of $Z_{cs}(3985)^-$ is assumed to favor $J^P=1^+$ and the quark content as $c\bar cs\bar u$ \cite{besiii2020Zcs}. It will be the first candidate of the hidden-charm four-quark state with strangeness.

Recall the theoretical investigations of the hidden-charm four-quark states with strangeness, the compact tetraquark configuration $sc\bar{q}\bar{c}$ has already been studied in the color-magnetic interaction method \cite{Cui2007} and QCD sum rules \cite{Wang:2009bd,Dias:2013qga,Di:2018dcf,Wang:2010rt,Wang:2009gx,Azizi:2020zyq,Sungu:2020zvk,Wang:2020iqt}. In Ref.\cite{PhysRevLett.110.232001}, the authors investigated the charged charmonium-like structures with hidden-charm and open-strange channels using the  initial single chiral particle emission mechanism. Their results suggested the existence of enhancement structures near the thresholds of $\bar{D}^{(*)}D_{s}^{(*)} $. In Ref.\cite{Lee:2008uy}, an axial-vector hidden-charm $D^{*-} D_{s}^{+}-D^{-} D_{s}^{*+}$ molecular state was also predicted to exist. 
Possible $D\bar{D}_{s0}^{*}(2317)$ and $D^{*}\bar{D}_{s1}^{*}(2460)$ molecules were studied in Ref.\cite{Di:2019qwv}, in which their results disfavor the existence of such states. 

A hadronic molecule is composed of two color-singlet hadrons by exchanging light mesons. This is a very useful configuration to study the nature of some exotic XYZ states and pentaquark states \cite{2016-Chen-p374-374,Chen:2016qju,2017-Lebed-p143-194,2018-Guo-p15004-15004,2019-Liu-p237-320,2020-Brambilla-p1-154}. 
Since the $Z_{cs}(3985)^-$ lie very close to the mass thresholds of $D_{s}^{-} D^{* 0}$ and $D_{s}^{*-} D^{0}$, it is naturally 
studied in a molecular picture \cite{1830582,1830580,1830608,1831047,1831062,Ozdem:2021yvo,Xu:2020evn,Yan:2021tcp}, as a partner state of $Z_c(3900)$ discovered by BESIII \cite{2013-Ablikim-p252001-252001}. It is also explained as a compound mixture of four different four-quark configurations \cite{1830632}, or a reflection structure of charmed-strange meson $D_{s2}^\ast(2573)$ \cite{1830623}. Besides, the production mechanisms of the hidden-charm four-quark states with strangeness are studied in Refs. \cite{1831033,1831054}. 
In Ref.\cite{Dias:2013qga}, the authors studied the decay width of the $D_s\bar D^\ast/D_s^\ast\bar D$ by calculating the three-point correlation functions in QCD sum rules. Their result of the total width suffers from a large uncertainty, although its central value is consistent with the experimental result of $Z_{cs}(3985)^-$. Such large uncertainty of the total width originated from the square of form factors, which is inherent and hard to be reduced in 
the method of three-point QCD sum rules. 
We also refers to the works \cite{Jin:2020yjn,Chen:2021uou,Shi:2021jyr,Guo:2020vmu,Ikeno:2021ptx,Meng:2021rdg,Simonov:2020ozp} for recent studies on $Z_{cs}(3985)$ in other methods. 
In this work, we shall study the exotic $\bar{D}_s^{(*)}D^{(*)}$ molecular states and $sc\bar q\bar c$ tetraquark states with $J^P=0^+, 1^+, 2^+$ in the method of QCD sum rules \cite{Shifman:1978bx,Reinders:1984sr,2000-Colangelo-p1495-1576}.

The paper is organized as follows. In Sec.~\Rmnum{2}, we construct the interpolating currents for the $\bar{D}_s^{(*)}D^{(*)}$ molecular systems and $sc\bar q\bar c$ tetraquark systems with $J^{P}=0^{+},1^{+}$ and $2^{+}$. In Sec.~\Rmnum{3}, we calculate the correlation functions and spectral densities for these interpolating currents. . We extract the masses for the $\bar{D}_s^{(*)}D^{(*)}$ molecular states and $sc\bar q\bar c$ tetraquark states by performing the QCD sum rule analyses in Sec.~\Rmnum{4}. The last section is a summary and discussion.

\section{Interpolating currents}
The color structures of a molecular field $[q \bar{Q}][Q \bar{q}]$ and a tetraquark field $[q Q][\bar Q \bar{q}]$ can be written via the SU(3) symmetry
\begin{equation}
\begin{aligned}
(\mathbf{3} \otimes \overline{\mathbf{3}})_{[q \bar{Q}]} \otimes(\mathbf{3} \otimes \overline{\mathbf{3}})_{[Q \bar{q}]} &=(\mathbf{1} \oplus \mathbf{8})_{[q \bar{Q}]} \otimes(\mathbf{1} \oplus \mathbf{8})_{[Q \bar{q}]} \\
&=(\mathbf{1} \otimes \mathbf{1}) \oplus(\mathbf{1} \otimes \mathbf{8}) \oplus(\mathbf{8} \otimes \mathbf{1}) \oplus(\mathbf{8} \otimes \mathbf{8}) \\
&=\mathbf{1} \oplus \mathbf{8} \oplus \mathbf{8} \oplus(\mathbf{1} \oplus \mathbf{8} \oplus \mathbf{8} \oplus \mathbf{1 0} \oplus \overline{\mathbf{1 0}} \oplus \mathbf{27})\, ,\\
(\mathbf{3} \otimes {\mathbf{3}})_{[q Q]} \otimes(\overline{\mathbf{3}} \otimes \overline{\mathbf{3}})_{[\bar Q \bar{q}]} &=(\mathbf{6} \oplus \overline{\mathbf{3}})_{[q Q]} \otimes(\mathbf{3} \oplus \overline{\mathbf{6}})_{[\bar Q \bar{q}]} \\
&=(\mathbf{6} \otimes \overline{\mathbf{6}}) \oplus(\overline{\mathbf{3}} \otimes \mathbf{3}) \oplus (\mathbf{6} \otimes \mathbf{3}) \oplus(\overline{\mathbf{3}} \otimes \overline{\mathbf{6}}) \\
&=(\mathbf{1} \oplus \mathbf{8} \oplus \mathbf{27}) \oplus(\mathbf{1} \oplus \mathbf{8}) \oplus (\mathbf{8} \oplus \mathbf{1 0}) \oplus (\mathbf{8} \oplus \overline{\mathbf{1 0}})\, ,
\label{SU3}
\end{aligned}
\end{equation}
in which the color singlet structures come from the $\left(\mathbf{1}_{[q \bar{Q}]} \otimes \mathbf{1}_{[Q \bar{q}]}\right)$ and $\left(\mathbf{8}_{[q \bar{Q}]} \otimes \mathbf{8}_{[Q \bar{q}]}\right)$ terms for the molecular field, while from the $\left(\mathbf{6}_{[q Q]} \otimes \overline{\mathbf{6}}_{[\bar Q \bar{q}]}\right)$ and $\left(\overline{\mathbf{3}}_{[q Q]} \otimes \mathbf{3}_{[\bar Q \bar{q}]}\right)$ terms for the tetraquark field. In this work, we shall consider the molecular and tetraquark interpolating currents with color structures $\left(\mathbf{1}_{[q \bar{Q}]} \otimes \mathbf{1}_{[Q \bar{q}]}\right)$ and $\left(\overline{\mathbf{3}}_{[q Q]} \otimes \mathbf{3}_{[\bar Q \bar{q}]}\right)$, respectively. To study the lowest lying molecular and tetraquark states, we use only S-wave mesonic and diquark fields to construct the molecular and tetraquark currents with the angular momentum $L=0$ between two mesonic fields and also two diquark fields. Finally, we obtain the $\bar{D}_s^{(*)}D^{(*)}$ molecular interpolating currents as 
\begin{equation}
\begin{aligned}
J_{1}&=(\bar{c}_{a} \gamma_{5} s_{a})(\bar{q}_{b} \gamma_{5} c_{b} )\, , ~~~~~~~ J^P=0^+\, , \\
J_{2}&=(\bar{c}_{a} \gamma_{\mu} s_{a})(\bar{q}_{b} \gamma^{\mu} c_{b} )\, , ~~~~~~~ J^P=0^+\, ,\\
J_{1\mu}&=(\bar{c}_{a} \gamma_{\mu} s_{a})(\bar{q}_{b}\gamma_{5} c_{b}   )\, ,~~~~~~~ J^P=1^+\, ,\\
J_{2\mu}&=(\bar{c}_{a} \gamma_{5} s_{a})(\bar{q}_{b}\gamma_{\mu} c_{b}   )\, ,~~~~~~~ J^P=1^+ \, ,\\
J_{3\mu}&=(\bar{c}_{a} \gamma^{\alpha} s_{a})(  \bar{q}_{b}\sigma_{\alpha\mu} \gamma_{5}c_{b} )\, ,~~ J^P=1^+\, ,\\
J_{4\mu}&=(\bar{c}_{a} \sigma_{\alpha\mu}\gamma_{5} s_{a})(  \bar{q}_{b} \gamma^{\alpha} c_{b} )\, , ~~ J^P=1^+\, ,\\
J_{\mu\nu}&= (\bar{c}_{a} \gamma_{\mu} s_{a})(  \bar{q}_{b}\gamma_{\nu}c_{b}  )\, ,~~~~~~~ J^P=2^+\, ,
\label{molecule_currents}
 \end{aligned}
\end{equation}
and the $sc\bar q\bar c$ tetraquark interpolating currents as 
\begin{equation}
\begin{aligned}
\eta_{1}&=s_{a}^{T} C \gamma_{5} c_{b}\left(\bar{q}_{a} \gamma_{5} C \bar{c}_{b}^{T}-\bar{q}_{b} \gamma_{5} C \bar{c}_{a}^{T}\right)\, , \, ~~~~~~~~~~~~ J^P=0^+\, , \\
\eta_{2}&=s_{a}^{T} C \gamma_{\mu} c_{b}\left(\bar{q}_{a} \gamma^{\mu} C \bar{c}_{b}^{T}-\bar{q}_{b} \gamma^{\mu} C \bar{c}_{a}^{T}\right)\, ,  \, ~~~~~~~~~~~ J^P=0^+\, ,\\
\eta_{1\mu}&=s_{a}^{T} C \gamma_{\mu} c_{b}\left(\bar{q}_{a} \gamma_{5} C \bar{c}_{b}^{T}-\bar{q}_{b} \gamma_{5} C \bar{c}_{a}^{T}\right)\, ,\, ~~~~~~~~~~~~ J^P=1^+\, ,\\
\eta_{2\mu}&=s_{a}^{T} C \gamma_{5} c_{b}\left(\bar{q}_{a} \gamma^{\mu} C \bar{c}_{b}^{T}-\bar{q}_{b} \gamma^{\mu} C \bar{c}_{a}^{T}\right)\, ,~~~~~~~~~~~~ J^P=1^+ \, ,\\
\eta_{3\mu}&=s_{a}^{T} C \gamma^{\alpha} c_{b}\left(\bar{q}_{a} \sigma_{\alpha\mu} \gamma_{5} C \bar{c}_{b}^{T}-\bar{q}_{b} \sigma_{\alpha\mu} \gamma_{5} C \bar{c}_{a}^{T}\right)\, ,\, ~~ J^P=1^+\, ,\\
\eta_{4\mu}&=s_{a}^{T} C \sigma_{\alpha\mu}\gamma_{5} c_{b}\left(\bar{q}_{a} \gamma^{\alpha} C \bar{c}_{b}^{T}-\bar{q}_{b} \gamma^{\alpha} C \bar{c}_{a}^{T}\right)\, ,~~~~~~~ J^P=1^+ \, ,\\
\eta_{\mu\nu}&= s_{a}^{T} C \gamma_{\mu} c_{b}\left(\bar{q}_{a} \gamma^{\nu} C \bar{c}_{b}^{T}-\bar{q}_{b} \gamma^{\nu} C \bar{c}_{a}^{T}\right)\, ,~~~~~~~~~~~~ J^P=2^+\, ,
\label{tetraquark_currents}
 \end{aligned}
\end{equation}
in which $a$, $b$ denote color indices and $q$ is an up or down quark. 
The mesonic field $\bar{q}_{a}\sigma_{\alpha\mu}\gamma_{5}q_{a}$ in $J_{3\mu}$ and $J_{4\mu}$ can couple to both the vector channel $J^{P}=1^{-}$($\bar{q}_{a}\sigma_{i j}\gamma_{5}q_{a}$) and axial-vector channel $J^{P}=1^{+}$ ($\bar{q}_{a}\sigma_{0 i}\gamma_{5}q_{a}$). We pick out its S-wave vector component by multiplicating a vector mesonic field $\bar{q}\gamma_{\alpha}q$, so that the molecular operators carry the positive parity. 
Similar situation happens for the tetraquark currents $\eta_{3\mu}$ and $\eta_{4\mu}$. The molecular currents in Eq. \eqref{molecule_currents} are not independent of the diquark-antidiquark currents in Eq. \eqref{tetraquark_currents}. Actually, a molecular current can be rewritten in terms of a sum over diquark-antidiquark currents via Fierz transformation with some suppression factors. In this work, we shall establish both the mass spectra for these two different configurations.
Using the interpolating currents in Eqs. \eqref{molecule_currents}-\eqref{tetraquark_currents}, we shall study the masses for the $\bar{D}_s^{(*)}D^{(*)}$ molecular states and $sc\bar q\bar c$ tetraquark states in the following. 

\section{QCD sum rules}
In this section, we study the two-point correlation functions of the scalar,  axial-vector and tensor interpolating currents above. For the scalar currents, the correlation function is
\begin{equation}
\begin{aligned}
\Pi\left(p^{2}\right)&=i \int d^{4} x e^{i p \cdot x}\left\langle 0\left|T\left[J(x) J^{\dagger}(0)\right]\right| 0\right\rangle\, ,
\end{aligned}
\end{equation}
and for the axial-vector current
\begin{equation}
\begin{aligned}
 \Pi_{\mu \nu}\left(p^{2}\right) =i \int d^{4} x e^{i p \cdot x}\left\langle 0\left|T\left[J_{\mu}(x) J_{\nu}^{\dagger}(0)\right]\right| 0\right\rangle\, . 
 \label{CF_AV}
\end{aligned}
\end{equation}
The correlation function $\Pi_{\mu\nu} (p^{2})$ in Eq.~\eqref{CF_AV} can be rewitten as 
\begin{equation}
\Pi_{\mu \nu}\left(p^{2}\right)=\left(\frac{p_{\mu} p_{\nu}}{p^{2}}-g_{\mu \nu}\right) \Pi_{1}\left(p^{2}\right)+\frac{p_{\mu} p_{\nu}}{p^{2}}\Pi_{0}\left(p^{2}\right)\, ,
\end{equation}
where $\Pi_{0}\left(p^{2}\right)$ and $\Pi_{1}\left(p^{2}\right)$ are the scalar and vector current polarization functions corresponding to the spin-0 and spin-1 intermediate states, respectively. The correlation function for the tensor current $J_{\mu\nu}(x)$ is
\begin{equation}
\begin{aligned}
 \Pi_{\mu \nu,\rho \sigma}\left(p^{2}\right) =i \int d^{4} x e^{i p \cdot x}\left\langle 0\left|T\left[J_{\mu\nu}(x) J_{\rho\sigma}^{\dagger}(0)\right]\right| 0\right\rangle\, ,
 \label{CF_T}
\end{aligned}
\end{equation}
which can be expressed as 
\begin{equation}
\Pi_{\mu \nu,\rho\sigma} \left(p^{2}\right)=\left(\eta_{\mu\rho}\eta_{\nu\sigma}+\eta_{\mu\sigma}\eta_{\nu\rho}-\frac{2}{3}\eta_{\mu\nu}\eta_{\rho\sigma}\right) \Pi_{2}\left(p^{2}\right)+\cdots \, ,
\end{equation}
where
\begin{equation}\
\eta_{\mu\nu}=\frac{p_{\mu} p_{\nu}}{p^{2}}-g_{\mu \nu},
\end{equation} 
and  $ \Pi_{2}\left(p^{2}\right)$ is the tensor current polarization functions related to the spin-2 intermediate states, and the $``\cdots"$ represents other spin-0 or spin-1 states.

At the hadronic level, the correlation function can be described via the dispersion relation
\begin{equation}
\Pi\left(p^{2}\right)=\frac{\left(p^{2}\right)^{N}}{\pi} \int_{4m_{c}^{2}}^{\infty} \frac{\operatorname{Im} \Pi(s)}{s^{N}\left(s-p^{2}-i \epsilon\right)} d s+\sum_{n=0}^{N-1} b_{n}\left(p^{2}\right)^{n}\, ,
\end{equation}
where $b_n$ is the subtraction constant. In QCD sum rules, the imaginary part of the correlation function is defined as the spectral function 
\begin{equation}
\rho (s)=\frac{1}{\pi} \text{Im}\Pi(s)=f_{H}^{2}\delta(s-m_{H}^{2})+\text{QCD continuum and higher states}\, ,
\end{equation}
in which the “pole plus continuum parametrization” is used. The parameters $f_{H}$ and $m_{H}$ are the coupling constant and mass of the lowest-lying hadronic resonance $H$ respectively 
\begin{equation}
\begin{aligned}
\langle 0|J| H\rangle &= f_{H}\, , \\
\left\langle 0\left|J_{\mu}\right| H\right\rangle &= f_{H} \epsilon_{\mu}\, , \\
\left\langle 0\left|J_{\mu\nu}\right| H\right\rangle &= f_{H} \epsilon_{\mu\nu}
 \end{aligned}
\end{equation}
with the polarization vector $\epsilon_{\mu}$ and polarization tensor $\epsilon_{\mu\nu}$.

On the other hand, we can calculate the correlation function $\Pi(p^{2})$ and spectral density $\rho(s)$ by means of operator product expansion (OPE) at the quark-gluon level. To evaluate the Wilson coefficients, we adopt the propagator of light quark in coordinate space and the propagator of heavy quark in momentum space
\begin{equation}
\begin{aligned}
i S_{q}^{a b}(x)=& \frac{i \delta^{a b}}{2 \pi^{2} x^{4}} \hat{x}
+\frac{i}{32 \pi^{2}} \frac{\lambda_{a b}^{n}}{2} g_{s} G_{\mu \nu}^{n} \frac{1}{x^{2}}\left(\sigma^{\mu \nu} \hat{x}+\hat{x} \sigma^{\mu \nu}\right)
-\frac{\delta^{a b} x^{2}}{12}\left\langle\bar{q} g_{s} \sigma \cdot G q\right\rangle
-\frac{m_{q} \delta^{a b}}{4 \pi^{2} x^{2}} \\
&+\frac{i \delta^{a b} m_{q}(\bar{q} q)}{48} \hat{x}
-\frac{i m_{q}\left\langle\bar{q} g_{s} \sigma \cdot G q\right) \delta^{a b} x^{2} \hat{x}}{1152}\, , \\
 i S_{Q}^{a b}(p)=& \frac{i \delta^{a b}}{\hat{p}-m_{Q}}
 +\frac{i}{4} g_{s} \frac{\lambda_{a b}^{n}}{2} G_{\mu \nu}^{n} \frac{\sigma^{\mu \nu}\left(\hat{p}+m_{Q}\right)+\left(\hat{p}+m_{Q}\right) \sigma^{\mu \nu}}{12}
 +\frac{i \delta^{a b}}{12}\left\langle g_{s}^{2} G G\right\rangle m_{Q} \frac{p^{2}+m_{Q} \hat{p}}{(p^{2}-m_{Q}^{2})^{4}}\, , 
 \end{aligned}
\end{equation}
where $q$ is $u$, $d$ or $s$ quark and $Q$ represents the $c$ or $b$ quark. The superscripts $a, b$ denote the color indices and $\hat{x}=x^{\mu}\gamma_{\mu},\hat{p}=p^{\mu}\gamma_{\mu}$. In this work, we calculate the Wilson coefficients up to dimension eight condensates at the leading order in $\alpha_s$. In Ref.~\cite{Albuquerque:2021tqd}, the NLO perturbative corrections to the correlation functions for the $sc\bar q\bar c$ tetraquark systems have been studied and their results show that such contributions are numerically small. The spectral densities for the interpolating currents in Eqs.~\eqref{molecule_currents}-\eqref{tetraquark_currents} are evaluated and listed in the appendix~\ref{spectral densities}. The tetraquark currents $\eta_{1}(x)$, $\eta_{2}(x)$, $\eta_{1\mu}(x)$, $\eta_{2\mu}(x)$ are the same with $\eta_{2}(x)$, $\eta_{4}(x)$, $\eta_{2\mu}(x)$, $\eta_{4\mu}(x)$ for the $sc\bar q\bar b$ systems in Ref. \cite{2020-Wang-p389-389}, by replacing the bottom quark to charm quark $b\to c$. Thus we don't list the spectral densities for these four tetraquark currents in the appendix~\ref{spectral densities}. To improve the convergence of the OPE series and suppress the contributions from continuum and higher states region, the Borel transformation is applied to the correlation function at both the hadron and the quark-gluon levels. The QCD sum rules are then established as
\begin{equation}
\mathcal{L}_{k}\left(s_{0}, M_{B}^{2}\right)=f_{H}^{2} m_{H}^{2 k} e^{-m_{H}^{2} / M_{B}^{2}}=\int_{4m_{c}^{2}}^{s_{0}} d s e^{-s / M_{B}^{2}} \rho(s) s^{k}\, ,
\label{Lk}
\end{equation}
in which $M_B$ represents the Borel mass introduced by the Borel transformation and $s_0$ is the continuum threshold.
The mass of the lowest-lying hadron can be thus extracted as
\begin{equation}
\begin{aligned}
m_{H}\left(s_{0}, M_{B}^{2}\right)=&\sqrt{\frac{\mathcal{L}_{1}\left(s_{0}, M_{B}^{2}\right)}{\mathcal{L}_{0}\left(s_{0}, M_{B}^{2}\right)}}\, ,
\end{aligned}
\end{equation}
which is the function of two parameters $M_B^2$ and $s_0$. We shall discuss the detail to obtain suitable parameter working regions in QCD sum rule analyses in next section. 

\section{Numerical analysis}
In this section, we perform the QCD sum rule analyses for the $\bar{D}_s^{(*)}D^{(*)}$ molecular and $sc\bar q\bar c$ tetraquark systems by using the interpolating currents in Eqs. \eqref{molecule_currents}-\eqref{tetraquark_currents}. We use the values of quark masses and various QCD condensates as follows \cite{Narison:1989aq,Jamin:2001zr,Jamin:1998ra,Ioffe:1981kw,Chung:1984gr,Dosch:1988vv,Khodjamirian:2011ub,Tanabashi:2018oca,PhysRevD.99.054505}
\begin{equation}
\begin{array}{l}
{m_{u}(2 \mathrm{GeV})=(2.2_{-0.4}^{+0.5} ) \mathrm{MeV}}\ , \vspace{1ex}  \\
{m_{d}(2 \mathrm{GeV})=(4.7_{-0.3}^{+0.5}) \mathrm{MeV}}\, ,\vspace{1ex}  \\
{m_{q}(2 \mathrm{GeV})=(3.5_{-0.2}^{+0.5}) \mathrm{MeV}}\, ,\vspace{1ex}  \\
{m_{s}(2 \mathrm{GeV})=(95_{-3}^{+9}) \mathrm{MeV}}\, ,\vspace{1ex}  \\
{m_{c}\left(m_{c}\right)=(1.275 _{-0.035}^{+0.025}) \mathrm{GeV}}\, , \vspace{1ex} \\
{m_{b}\left(m_{b}\right)=(4.18 _{-0.03}^{+0.04}) \mathrm{GeV}}\, , \vspace{1ex} \\
\langle\bar{q} q\rangle=-(0.24 \pm 0.03)^{3} \mathrm{GeV}^{3}\, , \vspace{1ex} \\
{\left\langle\bar{q} g_{s} \sigma \cdot G q\right\rangle=- M_{0}^{2}\langle\bar{q} q\rangle}\, ,\vspace{1ex}  \\
{ M_{0}^{2}=(0.8 \pm 0.2) \mathrm{GeV}^{2}}\, , \vspace{1ex} \\
{\langle\bar{s} s\rangle /\langle\bar{q} q\rangle= 0.8 \pm 0.1}\, , \vspace{1ex} \\
{\left\langle g_{s}^{2} G G\right\rangle= (0.48\pm0.14) \mathrm{GeV}^{4}}\, ,
\end{array}
\end{equation}
where the $u,d,s$ quark masses of  are the current quark masses obtained in the $\overline{MS}$ scheme at the scale $\mu = 2$ GeV.
We use the running mass in the $\overline{MS}$ scheme for the charm quark, which is different from the value of pole quark mass. 
Various literatures prove that the use of $\overline{MS}$ mass of the charm quark can lead to very good predictions for the masses of 
XYZ states in the framework of QCD sum rules~\cite{Chen:2016qju,Nielsen:2009uh}.

To establish a stable mass sum rule, one should find appropriate parameter working regions first, i.e, the continuum threshold $s_{0}$ and the Borel mass $M_{B}^{2}$. The threshold $s_{0}$ can be determined via the minimized variation of the hadronic mass $m_{H}$ with the Borel mass $M_{B}^{2}$. The lower bound on Borel mass $M_{B}^{2}$ can be fixed by requiring a reasonable OPE convergence while its upper bound is determined through a sufficient pole contribution. The pole contribution is defined as
\begin{equation}
\mathrm{PC}\left(s_{0}, M_{B}^{2}\right)=\frac{\mathcal{L}_{0}\left(s_{0}, M_{B}^{2}\right)}{\mathcal{L}_{0}\left(\infty, M_{B}^{2}\right)}\, ,
\end{equation}
where $\mathcal{L}_{0}$ has been defined in Eq.~(\ref{Lk}). 

\begin{figure}[h]
\centering
\includegraphics[width=10cm]{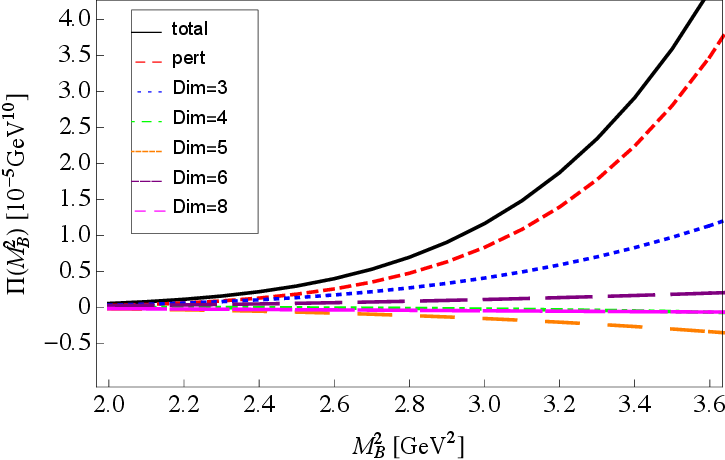}\\
\caption{OPE convergence for the $\bar{D}_s^\ast D^\ast$ molecular current $J_{2}(x)$ with $J^{P}=0^+$.}
\label{FigratioJ1}
\end{figure}

We use the $\bar{D}_s^\ast D^\ast$ molecular current $J_{2}(x)$ with $J^{P}=0^+$ as an example to show the detail of the numerical analysis. For this current, the dominant non-perturbative contribution to the correlation function comes from the quark condensate $\langle\bar{q}q\rangle$ and $\langle\bar{s}s\rangle$.  In Fig.~\ref{FigratioJ1}, we show the contributions of the perturbative term and various condensate terms to the correlation function. It is clear that the Borel mass $M_{B}^{2}$ should be large enough to ensure the convergence of OPE series. Here, we require that the highest dimension condensate contribution to be less than 10\%, 
\begin{equation}
\begin{array}{c}
{\frac{\Pi^{\langle\bar{q}q\rangle \langle\bar{q}g_{s}\sigma\cdot G q\rangle}(M_{B}^{2},\infty)}{\Pi(M_{B}^{2},\infty)}<10\% } \, ,
\end{array}
\end{equation}
which results in $M_{B}^{2}\geq 2.6\text{GeV}^{2}$.

\begin{figure}[h]
\centering
  \includegraphics[width=8.5cm]{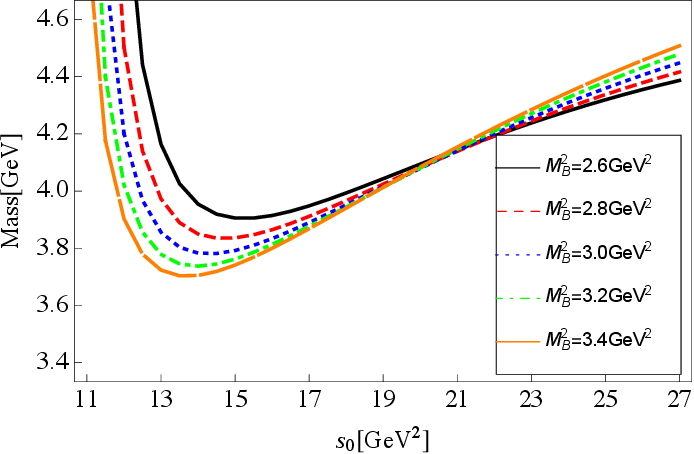}\quad
  \includegraphics[width=8.5cm]{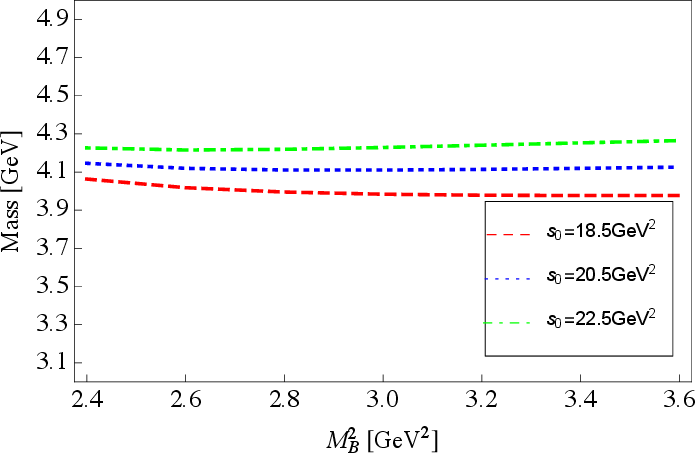}\\
\caption{Variations of $m_{H}$ with $s_{0}$ and $M_{B}^{2}$ corresponding to the current $J_{2}(x)$ in the $\bar{D}_s^\ast D^\ast$ system with $J^{P}= 0^{+}$.}
  \label{FigMBandstaJ1}
\end{figure}

As mentioned above, the variation of the output hadron mass $m_{H}$ with $M_{B}^{2}$ should be minimized to obtain the optimized value of the continuum threshold $s_0$. We show the variations of $m_{H}$ with $s_{0}$ and $M_{B}^{2}$ in Fig.~\ref{FigMBandstaJ1},  from which the dependence of $m_{H}$ on $M_{B}^{2}$ can be minimized around $s_{0}\approx 20.5\text{GeV}^{2}$. Requiring the pole contribution be larger than 30\%, the upper bound on $M_{B}^{2}$ can then be determined to be $3.4~\text{GeV}^{2}$. The working region of the Borel parameter for the scalar $\bar{D}_s^\ast D^\ast$ molecular current $J_{2}(x)$ is thus $2.6\leq M_{B}^{2}\leq 3.4\text{GeV}^{2}$. As shown in Fig.~\ref{FigMBandstaJ1}, the mass sum rules are established to be very stable in these parameter regions and the hadron mass for the $\bar{D}_s^\ast D^\ast$ molecule with $J^{P}= 0^{+}$ can be obtained as 
\begin{align}
m_{\bar{D}_s^\ast D^\ast, \, 0^+}=4.11\pm0.14 \text{GeV}\, ,
\end{align}
in which the error comes from the uncertainties of the continuum threshold $s_{0}$, Borel mass $M_B$, the various condensates and quark masses.  
After performing similar analyses, we obtain the numerical results for all the other interpolating currents  in Eqs.~\eqref{molecule_currents}-\eqref{tetraquark_currents} and collect them in Table~\ref{resultone}. 
\begin{table}[h!]
\renewcommand{\arraystretch}{1.4}
\caption{The numerical results for the $\bar{D}_s^{(\ast)} D^{(\ast)}$ molecular and diquark-antiquark $sc\bar q\bar c$ tetraquark systems.}
\begin{center}
\label{resultone}
\begin{ruledtabular}
\begin{tabular}{ccccccc }
System &   Current&  $J^{P} $  &$s_{0}$(\text{GeV$^{2}$}) & $M_{B}^{2}$(\text{GeV$^{2}$})  &$ m_{H}$(\text{GeV})  &PC(\%)  \\  \hline
 $\bar{D}_sD$ &    $J_{1}$ &     $0^{+}$  &   18.0 $\pm$ 2.0              & 1.6 $\sim$ 3 .6         & 3.74 $\pm$ 0.13  &52.5    \\ 
 $\bar{D}_s^\ast D^\ast$ &     $J_{2}$ &     $0^{+}$  &   20.5 $\pm$ 2.0              & 2.6 $\sim$ 3.4         & 4.11 $\pm$ 0.14  &42.4     
 \vspace{.2cm}\\
$\bar{D}_s^{*}D$ &     $J_{1\mu}$ &     $1^{+}$  &   20.7 $\pm$ 2.0           & 2.1 $\sim$ 2.5      & 3.99 $\pm$ 0.12  &68.2    \\
$\bar{D}_sD^{*}$ &     $J_{2\mu}$ &     $1^{+}$  &   20.5 $\pm$ 2.0              & 2.1 $\sim$ 2.5      & 3.97 $\pm$ 0.11  &67.7   \\
$\bar{D}_s^\ast D^\ast$ &  $J_{3\mu}$ &      $1^{+}$  &   21.5 $\pm$ 2.0         & 2.8 $\sim$ 3.6             & 4.22 $\pm$ 0.14  & 40.1      \\  
$\bar{D}_s^\ast D^\ast$ &  $J_{4\mu}$ &      $1^{+}$  &   21.5 $\pm$ 2.0           & 2.8 $\sim$ 3.6            & 4.22 $\pm$ 0.14   &40.0   \vspace{.2cm}   \\ 
$\bar{D}_s^\ast D^\ast$ & $J_{\mu\nu}$ &     $2^{+}$  &    23.0 $\pm$ 2.0            & 2.8 $\sim$ 4.3           & 4.34 $\pm$ 0.13   &48.7     \\
\hline
$\mathbf{0}_{[sc]} \oplus \mathbf{0}_{[\bar q \bar{c}]}$ (spin-spin) &    $\eta_{1}$ &     $0^{+}$  &   18.0 $\pm$ 2.0              & 2.1 $\sim$ 3 .1         & 3.84 $\pm$ 0.15  &46.3    \\ 
$\mathbf{1}_{[sc]} \oplus \mathbf{1}_{[\bar q \bar{c}]}$ &     $\eta_{2}$ &     $0^{+}$  &   20.0 $\pm$ 2.0              & 2.6 $\sim$ 3.2        & 4.13 $\pm$ 0.17  & 35.6     
 \vspace{.2cm}\\
$\mathbf{1}_{[sc]} \oplus \mathbf{0}_{[\bar q \bar{c}]}$ &     $\eta_{1\mu}$ &     $1^{+}$  &  19.0 $\pm$ 2.0           & 2.5 $\sim$ 3.3      & 3.98 $\pm$ 0.16  & 41.0    \\
$\mathbf{0}_{[sc]} \oplus \mathbf{1}_{[\bar q \bar{c}]}$ &     $\eta_{2\mu}$ &     $1^{+}$  &  19.0 $\pm$ 2.0              & 2.5 $\sim$ 3.3      & 3.97 $\pm$ 0.15  &41.6   \\
$\mathbf{1}_{[sc]} \oplus \mathbf{1}_{[\bar q \bar{c}]}$ &  $\eta_{3\mu}$ &      $1^{+}$     &  22.0 $\pm$ 2.0         & 2.9 $\sim$ 3.6      & 4.28 $\pm$ 0.14  & 40.9      \\  
$\mathbf{1}_{[sc]} \oplus \mathbf{1}_{[\bar q \bar{c}]}$ &  $\eta_{4\mu}$ &      $1^{+}$  &   22.0 $\pm$ 2.0           & 2.9 $\sim$ 3.6            & 4.28 $\pm$ 0.14   &41.1   \vspace{.2cm}   \\ 
$\mathbf{1}_{[sc]} \oplus \mathbf{1}_{[\bar q \bar{c}]}$ & $\eta_{\mu\nu}$ &     $2^{+}$  &    23.0 $\pm$ 2.0            & 2.8 $\sim$ 4.3           & 4.33 $\pm$ 0.13   &46.4     \\
\end{tabular}
\end{ruledtabular}
\end{center}
\end{table}

In Table~\ref{resultone}, the mass of scalar $\bar{D}_sD$ molecular state is predicted to be slightly below the open-charm threshold $T_{\bar{D}_sD}=3.84$ GeV, implying that it can only decay into the hidden-charm channel $\eta_c K$. The scalar $\bar{D}_s^{*} D^{*}$ state is predicted to be very close to $T_{\bar{D}_s^{*} D^{*}}=4.12$ GeV, however, it can decay into $\bar{D}_sD$ and $\eta_c K$ final states kinematically in S-wave. The masses for the  
$\bar{D}_{s}^{*} D^{*}$ molecular states with $J^P=1^{+}, 2^{+}$ are significantly above the  corresponding open-charm thresholds. 

The masses obtained from the axial-vector molecular currents $J_{1\mu}$ and $J_{2\mu}$ are $m_{\bar{D}_s^{*}D, \, 1^+}=(3.99 \pm 0.12)$ GeV, $m_{\bar{D}_{s}D^{*}, \, 1^+}=(3.97 \pm 0.11)$ GeV, which are almost degenerate with each other. One may wonder whether these two currents $J_{1\mu}$ and $J_{2\mu}$ could couple to the same physical molecular state or not. In QCD sum rules, this can be specified by studying the following 
off-diagonal correlation function
\begin{equation}
\begin{aligned}
 \Pi_{12\mu \nu}^M\left(p^{2}\right) =i \int d^{4} x e^{i p \cdot x}\left\langle 0\left|T\left[J_{1\mu}(x) J_{2\nu}^{\dagger}(0)\right]\right| 0\right\rangle\, . 
 \label{CF_off_molecule}
\end{aligned}
\end{equation}
Our calculation shows that this off-diagonal correlation function $\Pi_{12\mu \nu}^M\left(p^{2}\right)=0$ at the leading order of $\alpha_s$ for the axial-vector molecular currents $J_{1\mu}$ and $J_{2\mu}$, including the perturbative term and all contributions from various non-perturbative condensates. 
According to Ref. \cite{Albuquerque:2021tqd}, the NLO perturbative correction is numerically small and thus $\Pi_{12\mu \nu}^M\left(p^{2}\right)$ is still negligible comparing to the diagonal correlators $\Pi_{11\mu \nu}^M\left(p^{2}\right)$ and $\Pi_{22\mu \nu}^M\left(p^{2}\right)$ at the next leading order of $\alpha_s$. Such a result implies that $J_{1\mu}$ and $J_{2\mu}$ may couple to different physical states. 

\begin{figure}[h]
\centering
  \includegraphics[width=8.5cm]{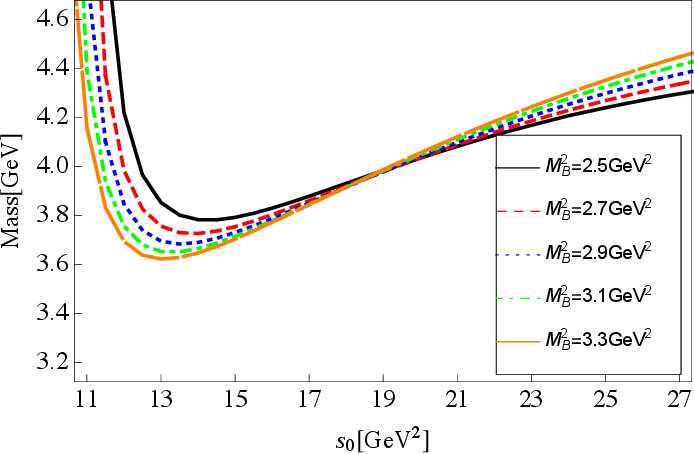}\quad
  \includegraphics[width=8.5cm]{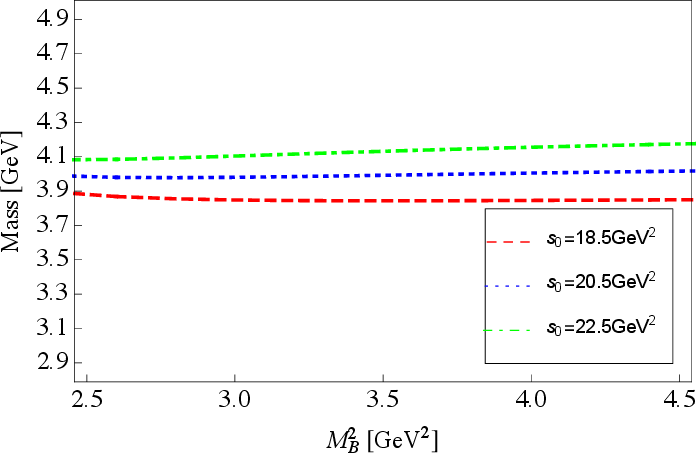}\\
\caption{Variations of $m_{H}$ with $s_{0}$ and $M_{B}^{2}$ for the current $\eta_{1\mu}(x)$ in the $sc\bar q\bar c$ tetraquark system with $J^{P}= 1^{+}$.}
  \label{FigMBandsta-eta1}
\end{figure}

We also study the $sc\bar q\bar c$ tetraquark systems with $J^P=0^+, 1^+, 2^+$. In Fig. \ref{FigMBandsta-eta1}, we show the variations of the tetraquark mass with $s_{0}$ and $M_{B}^{2}$ for the current $\eta_{1\mu}(x)$ with $J^{P}= 1^{+}$, and the mass sum rules are very stable and reliable at the chosen parameter regions. For the interpolating currents in Eq. \eqref{tetraquark_currents},  we collect the numerical results for these $sc\bar q\bar c$ tetraquark systems in Table~\ref{resultone}. It is shown that the mass spectra for the $sc\bar q\bar c$ tetraquarks are very similar with the $\bar{D}_s^{(\ast)} D^{(\ast)}$ molecular states. For the axial-vector $sc\bar q\bar c$ tetraquark systems, the extracted masses from $\eta_{1\mu}(x)$ and $\eta_{2\mu}(x)$ are almost the same with the $\bar{D}_s^{*}D$ and $\bar{D}_sD^{*}$ molecular states, which are consistent with the mass of $Z_{cs}(3985)^-$ from BESIII \cite{besiii2020Zcs}. 
It is interesting to examine the off-diagonal correlation function for $\eta_{1\mu}(x)$ and $\eta_{2\mu}(x)$
\begin{equation}
\begin{aligned}
 \Pi_{12\mu \nu}^T\left(p^{2}\right) =i \int d^{4} x e^{i p \cdot x}\left\langle 0\left|T\left[J_{1\mu}(x) J_{2\nu}^{\dagger}(0)\right]\right| 0\right\rangle\, .
 \label{CF_off_tetraquark}
\end{aligned}
\end{equation}
The calculation indicates that the perturbative term and the quark condensate terms in $\Pi_{12\mu \nu}^T\left(p^{2}\right)$ are equal to zero, 
This off-diagonal correlation function $\Pi_{12\mu \nu}^T\left(p^{2}\right)$ is very small, suggesting that the currents $\eta_{1\mu}(x)$ and $\eta_{2\mu}(x)$ cannot strongly couple to the same physical state.

\section{Conclusion}
To study the hidden-charm four-quark systems with strangeness, we have calculated the mass spectra for the $\bar{D}_s^{(*)}D^{(*)}$ molecular states and $sc\bar q\bar c$ tetraquark states with $J^P=0^+, 1^+, 2^+$ in the framework of QCD sum rules. We construct the corresponding molecular and tetraquark interpolating currents, calculate their two-point correlation functions and spectral densities up to dimension eight condensates at the leading order of $\alpha_s$. The quark condensates are found to be the most important non-perturbative contribution to the correlation functions for both molecular and tetraquark systems. 

One may wonder if the two-meson scattering states can contribute to the correlation functions in our calculations. 
In general, the interpolating currents can couple to all structures with the same quantum numbers, including resonances, two-meson scattering states 
and continuum. And thus these structures will give contributions to the correlation functions. However, it has been demonstrated that the two-meson scattering states cannot saturate the QCD sum rules, while only exotic four-quark states can saturate the QCD sum rules. Moreover, the contributions from the two-meson scattering states to the correlation functions are numerically negligible~\cite{Wang:2020cme,Albuquerque:2021tqd}.

Our results show that the masses of the axial-vector $\bar{D}_sD^{*}$, $\bar{D}_s^{*}D$ molecular states and the $sc\bar q\bar c$ tetraquark states from $\eta_{1\mu}$, $\eta_{2\mu}$ are calculated in good agreement with the mass of $Z_{cs}(3985)^-$. The present calculations are difficult for distinguishing the nature of $Z_{cs}(3985)^-$ from the molecular and diquark-antidiquark configurations. In both the molecular and diquark-antidiquark pictures, our results suggest that there may exist two almost degenerate states, as the strange partners of the $X(3872)$ and $Z_c(3900)$. We propose to carefully examine the $Z_{cs}(3985)$ in future experiments to verify this. One can also search for more hidden-charm four-quark states with strangeness not only in the open-charm $\bar{D}_s^{(*)}D^{(*)}$ channels, but also in the hidden-charm channels $\eta_c K/K^\ast$, $J/\psi K/K^\ast$. 

{\bf\emph{Note added:}} After we finished this work, the LHCb Collaboration has reported two new charged resonances $Z_{cs}(4000)^+$ and $Z_{cs}(4220)^+$ in the $J/\psi K^+$ final states~\cite{Aaij:2021ivw}. Their masses and decay widths are measured as $M_{Z_{cs}(4000)^+}=4003\pm6^{+4}_{-14}$ MeV, $\Gamma_{Z_{cs}(4000)^+}=131\pm15\pm26$ MeV and $M_{Z_{cs}(4220)^+}=4216\pm24^{+43}_{-30}$ MeV, $\Gamma_{Z_{cs}(4220)^+} =233\pm52^{+97}_{-73}$ MeV, while their spin-parity quantum numbers are identified to prefer $J^P=1^+$. These masses and spin-parity are consistent with the axial-vector $\bar{D}_sD^{*}$ ($\bar{D}_s^{*}D$), $\bar{D}^{*}_sD^{*}$  molecular states and $\mathbf{1}_{[sc]} \oplus \mathbf{0}_{[\bar q \bar{c}]}$ ($\mathbf{0}_{[sc]} \oplus \mathbf{1}_{[\bar q \bar{c}]}$), $\mathbf{1}_{[sc]} \oplus \mathbf{1}_{[\bar q \bar{c}]}$ ($\mathbf{1}_{[sc]} \oplus \mathbf{1}_{[\bar q \bar{c}]}$) tetraquark states that we have predicted in Table~\ref{resultone}.

According to LHCb's observation, the decay width of $Z_{cs}(4000)$ is much larger than that of $Z_{cs}(3985)$ observed by BESIII~\cite{besiii2020Zcs}. LHCb found no evidence that the $Z_{cs}(4000)$ and $Z_{cs}(3985)$ are the same state, although their masses are very close to each other. It this is true, they may be identified as the strange partners of the $X(3872)$ and $Z_c(3900)$ with $J^{PC}=1^{++}$ and $J^{PC}=1^{+-}$ respectively. We propose to carefully examine the $Z_{cs}(4000)$ and $Z_{cs}(3985)$ in future experiments to verify this.

\section*{ACKNOWLEDGMENTS}

This work is supported in part by National Key R$\&$D Program of China under Contracts No. 2020YFA0406400, the National Natural Science Foundation of China under Grants No. 11722540 and No. 12075019, the Fundamental Research Funds for the Central Universities.

\appendix
 \section{The spectral densities  \label{spectral densities}}
In this appendix, we list the spectral densities for the $\bar{D}_s^{(*)}D^{(*)}$ and $sc\bar q\bar c$ systems with $J^{P}=0^{+}$, $1^{+}$ and $2^{+}$. The spectral density includes the perturbative term, quark condensate, gluon condensate, quark-gluon mixed condensate, four-quark condensate  and dimension eight condensate  
\begin{equation}
\begin{aligned}
\rho(s)=\rho^{0}(s)+\rho^{3}(s)+\rho^{4}(s)+\rho^{5}(s)+\rho^{6}(s)+\rho^{8}(s)\, ,
\end{aligned}
\end{equation}
in which the superscripts stand for the dimension of various condensates.

1. Spectral densities for $J_{1}$:\\
\begin{equation}\nonumber
\rho_{J_{1}}^{0a}(s)=
\frac{3} {2048 \pi^{6}}\int_{\alpha_{min}}^{\alpha_{max}} d \alpha \int_{\beta_{min}}^{\beta_{max}} d \beta
 \frac{(1-\alpha-\beta)^{2}} { \alpha^{3} \beta^{3}}(m_{c}^{2}(\alpha+\beta)-\alpha \beta s)^{3}(m_{c}^{2}(\alpha+\beta)-3 \alpha \beta s)\, ,
\end{equation}

\begin{equation}\nonumber
\rho_{J_{1}}^{0b}(s)=
- \frac{3m_{c}} {1024 \pi^{6}}\int_{\alpha_{min}}^{\alpha_{max}} d \alpha \int_{\beta_{min}}^{\beta_{max}} d \beta
(1-\alpha-\beta)^{2}(m_{c}^{2}(\alpha+\beta)-\alpha \beta s)^{2}(2m_{c}^{2}(\alpha+\beta)-5 \alpha \beta s) \Big(\frac{m_{s}}{\alpha^{2} \beta^{3}}+\frac{m_{q}}{\alpha^{3} \beta^{2}}\Big)\, ,
\end{equation}

 \begin{equation}\nonumber
 \begin{aligned}
\rho_{J_{1}}^{3a}(s)=&
- \frac{3\langle\bar{s}s\rangle}{128\pi^{4} }
\int_{\alpha_{min}}^{\alpha_{max}} d \alpha \int_{\beta_{min}}^{\beta_{max}} d \beta
(m_{c}^{2}(\alpha+\beta)-\alpha \beta s)\Big[\frac{2(1-\alpha-\beta)(m_{c}^{2}(\alpha+\beta)-2\alpha \beta s)m_{c}} { \alpha \beta^{2}}\\
 &-\frac{2m_{c}^{2}m_{q}+(m_{c}^{2}(\alpha+\beta)-2\alpha \beta s)m_{s}}{\alpha\beta}\Big]\, ,
\end{aligned}
\end{equation}

 \begin{equation}\nonumber
 \begin{aligned}
\rho_{J_{1}}^{3b}(s)=&
- \frac{3\langle\bar{q}q\rangle}{128\pi^{4} }
\int_{\alpha_{min}}^{\alpha_{max}} d \alpha \int_{\beta_{min}}^{\beta_{max}} d \beta
(m_{c}^{2}(\alpha+\beta)-\alpha \beta s)\Big[\frac{2(1-\alpha-\beta)(m_{c}^{2}(\alpha+\beta)-2\alpha \beta s)m_{c}} { \alpha^{2} \beta}\\
 &-\frac{2m_{c}^{2}m_{s}+(m_{c}^{2}(\alpha+\beta)-2\alpha \beta s)m_{q}}{\alpha\beta}\Big]\, ,
\end{aligned}
\end{equation}

\begin{equation}\nonumber
\begin{aligned}
\rho_{J_{1}}^{4a}(s)=
&\frac{\langle g_{s}^{2} G G\rangle m_{c}^{2} }{4096 \pi^{6}}\int_{\alpha_{min}}^{\alpha_{max}} d \alpha \int_{\beta_{min}}^{\beta_{max}} d \beta (1-\alpha-\beta)^{2}
(2m_{c}^{2}(\alpha+ \beta)-3 \alpha \beta s)\Big(\frac{1}{\alpha^{3}}+\frac{1}{\beta^{3}}\Big)\, ,
\end{aligned}
\end{equation}

\begin{equation}\nonumber
\begin{aligned}
\rho_{J_{1}}^{4b}(s)=
&\frac{3\langle g_{s}^{2} G G\rangle m_{c}^{2} }{2048 \pi^{6}}\int_{\alpha_{min}}^{\alpha_{max}} d \alpha \int_{\beta_{min}}^{\beta_{max}} d \beta (1-\alpha-\beta)
(m_{c}^{2}(\alpha+ \beta)-\alpha \beta s)(m_{c}^{2}(\alpha+ \beta)-2\alpha \beta s)\Big(\frac{1}{\alpha^{2}\beta}+\frac{1}{\alpha\beta^{2}}\Big)\, ,
\end{aligned}
\end{equation}

\begin{equation}\nonumber
\begin{aligned}
\rho_{J_{1}}^{5a}(s)=
&\frac{3 \langle\bar{s} g_{s}\sigma\cdot G s\rangle m_{c}}{256 \pi^{4}} \int_{\alpha_{min}}^{\alpha_{max}} d \alpha \int_{\beta_{min}}^{\beta_{max}} d \beta
\Big[(2 m_{c}^{2}(\alpha+\beta)-3 s \alpha \beta) \Big(\frac{1}{\beta}-\frac{2(1-\alpha-\beta)}{\beta^{2}}\Big)+\frac{2m_{c}m_{q}}{\beta}\Big]\, ,
\end{aligned}
\end{equation}

\begin{equation}\nonumber
\begin{aligned}
\rho_{J_{1}}^{5b}(s)=
&\frac{3 \langle\bar{q} g_{s}\sigma\cdot G q\rangle m_{c}}{256 \pi^{4}} \int_{\alpha_{min}}^{\alpha_{max}} d \alpha \int_{\beta_{min}}^{\beta_{max}} d \beta
 \Big[(2 m_{c}^{2}(\alpha+\beta)-3 s \alpha \beta) \Big(\frac{1}{\alpha}-\frac{2(1-\alpha-\beta)}{\alpha^{2}}\Big)+\frac{2m_{c}m_{s}}{\alpha}\Big]\, ,
\end{aligned}
\end{equation}

\begin{equation}\nonumber
\begin{aligned}
\rho_{J_{1}}^{5c}(s)=
&\frac{\langle\bar{q} g_{s}\sigma\cdot G q\rangle}{512 \pi^{4}}\left(\left(s-2 m_{c}^{2}\right) m_{q}-6 m_{c}^{2} m_{s}\right) \sqrt{1-\frac{4 m_{c}^{2}}{s}}\\
+&\frac{\langle\bar{s} g_{s}\sigma\cdot G s\rangle}{512 \pi^{4}}\left(\left(s-2 m_{c}^{2}\right) m_{s}-6 m_{c}^{2} m_{q}\right) \sqrt{1-\frac{4 m_{c}^{2}}{s}}\, ,
\end{aligned}
\end{equation}

\begin{equation}\nonumber
\rho_{J_{1}}^{6a}(s)=
\frac{\langle\bar{s}s\rangle\langle\bar{q}q\rangle }{32 \pi^{2}}(2m_{c}^{2}+m_{c}m_{q}+m_{c}m_{s})
\sqrt{1-\frac{4 m_{c}^{2}}{s}}\, ,
\end{equation}

\begin{equation}\nonumber
\Pi_{J_{1}}^{6b}\left(M_{B}^{2}\right)=
-\frac{\langle\bar{s}s\rangle\langle\bar{q}q\rangle m_{c}^{3}}{32 \pi^{2}} \int_{0}^{1} d \alpha\Big(\frac{m_{q}}{1-\alpha}+\frac{m_{s}}{\alpha}\Big) e^{\frac{-m_{c}^{2}} { \alpha(1-\alpha) M_{B}^{2}}}\, ,
\end{equation}

\begin{equation}\nonumber
\Pi_{J_{1}}^{8}\left(M_{B}^{2}\right)=
\frac{ m_{c}^{4}}{64 \pi^{2}} \int_{0}^{1} d \alpha\Big(\frac{ \langle\bar{s}s\rangle \langle\bar{q}g_{s}\sigma\cdot Gq\rangle+\langle\bar{q}q\rangle \langle\bar{s}g_{s}\sigma\cdot Gs\rangle}{(1-\alpha)^{2}M_{B}^{2}}
-\frac{2 \langle\bar{s}s\rangle \langle\bar{q}g_{s}\sigma\cdot Gq\rangle}{(1-\alpha)m_{c}^{2}}-\frac{ 2\langle\bar{q}q\rangle \langle\bar{s}g_{s}\sigma\cdot Gs\rangle}{\alpha m_{c}^{2}}\Big) e^{\frac{-m_{c}^{2}} { \alpha(1-\alpha) M_{B}^{2}}}\, ,
\end{equation}
where
\begin{equation}\nonumber
\begin{aligned}
\alpha_{min }&=\frac{1}{2}-\frac{1}{2}\sqrt{1-\frac{4 m_{c}^{2}}{s}},~~~
\alpha_{max }=\frac{1}{2}+\frac{1}{2}\sqrt{1-\frac{4 m_{c}^{2}}{s}},~~~
\beta_{min}=\frac{\alpha m_{c}^{2}}{\alpha s-m_{c}^{2}},~~~~\beta_{max}=1-\alpha \, ,
 \end{aligned}
\end{equation}

2. Spectral densities for $J_{2}$:\\
\begin{equation}\nonumber
\rho_{J_{2}}^{0a}(s)=
\frac{3} {512 \pi^{6}}\int_{\alpha_{min}}^{\alpha_{max}} d \alpha \int_{\beta_{min}}^{\beta_{max}} d \beta
 \frac{(1-\alpha-\beta)^{2}} { \alpha^{3} \beta^{3}}(m_{c}^{2}(\alpha+\beta)-\alpha \beta s)^{3}(m_{c}^{2}(\alpha+\beta)-3 \alpha \beta s)\, ,
\end{equation}

\begin{equation}\nonumber
\rho_{J_{2}}^{0b}(s)=
- \frac{3m_{c}} {512 \pi^{6}}\int_{\alpha_{min}}^{\alpha_{max}} d \alpha \int_{\beta_{min}}^{\beta_{max}} d \beta
(1-\alpha-\beta)^{2}(m_{c}^{2}(\alpha+\beta)-\alpha \beta s)^{2}(2m_{c}^{2}(\alpha+\beta)-5 \alpha \beta s) \Big(\frac{m_{s}}{\alpha^{2} \beta^{3}}+\frac{m_{q}}{\alpha^{3} \beta^{2}}\Big)\, ,
\end{equation}

 \begin{equation}\nonumber
 \begin{aligned}
\rho_{J_{2}}^{3a}(s)=&
- \frac{3\langle\bar{s}s\rangle}{64\pi^{4} }
\int_{\alpha_{min}}^{\alpha_{max}} d \alpha \int_{\beta_{min}}^{\beta_{max}} d \beta
(m_{c}^{2}(\alpha+\beta)-\alpha \beta s)\Big[\frac{2(1-\alpha-\beta)(m_{c}^{2}(\alpha+\beta)-2\alpha \beta s)m_{c}} { \alpha \beta^{2}}\\
 &-\frac{4m_{c}^{2}m_{q}+(m_{c}^{2}(\alpha+\beta)+2\alpha \beta s)m_{s}}{\alpha\beta}\Big]\, ,
\end{aligned}
\end{equation}

 \begin{equation}\nonumber
 \begin{aligned}
\rho_{J_{2}}^{3b}(s)=&
- \frac{3\langle\bar{q}q\rangle}{64\pi^{4} }
\int_{\alpha_{min}}^{\alpha_{max}} d \alpha \int_{\beta_{min}}^{\beta_{max}} d \beta
(m_{c}^{2}(\alpha+\beta)-\alpha \beta s)\Big[\frac{2(1-\alpha-\beta)(m_{c}^{2}(\alpha+\beta)-2\alpha \beta s)m_{c}} { \alpha^{2} \beta}\\
 &-\frac{4m_{c}^{2}m_{s}+(m_{c}^{2}(\alpha+\beta)+2\alpha \beta s)m_{q}}{\alpha\beta}\Big]\, ,
\end{aligned}
\end{equation}

\begin{equation}\nonumber
\begin{aligned}
\rho_{J_{2}}^{4}(s)=
&\frac{\langle g_{s}^{2} G G\rangle m_{c}^{2} }{1024 \pi^{6}}\int_{\alpha_{min}}^{\alpha_{max}} d \alpha \int_{\beta_{min}}^{\beta_{max}} d \beta (1-\alpha-\beta)^{2}
(2m_{c}^{2}(\alpha+ \beta)-3 \alpha \beta s)\Big(\frac{1}{\alpha^{3}}+\frac{1}{\beta^{3}}\Big)\, ,
\end{aligned}
\end{equation}

\begin{equation}\nonumber
\begin{aligned}
\rho_{J_{2}}^{5a}(s)=
&\frac{3 \langle\bar{s} g_{s}\sigma\cdot G s\rangle m_{c}}{128 \pi^{4}} \int_{\alpha_{min}}^{\alpha_{max}} d \alpha \int_{\beta_{min}}^{\beta_{max}} d \beta
\frac{2 m_{c}^{2}(\alpha+\beta)-3 s \alpha \beta}{\beta}\\
 &+\frac{3 \langle\bar{q} g_{s}\sigma\cdot G q\rangle m_{c}}{128 \pi^{4}} \int_{\alpha_{min}}^{\alpha_{max}} d \alpha \int_{\beta_{min}}^{\beta_{max}} d \beta
\frac{2 m_{c}^{2}(\alpha+\beta)-3 s \alpha \beta }{\alpha}\, ,
\end{aligned}
\end{equation}

\begin{equation}\nonumber
\begin{aligned}
\rho_{J_{2}}^{5b}(s)=
&\frac{\langle\bar{q} g_{s}\sigma\cdot G q\rangle}{128 \pi^{4}}\left(\left(s-2 m_{c}^{2}\right) m_{q}-6 m_{c}^{2} m_{s}\right) \sqrt{1-\frac{4 m_{c}^{2}}{s}}\\
+&\frac{\langle\bar{s} g_{s}\sigma\cdot G s\rangle}{128 \pi^{4}}\left(\left(s-2 m_{c}^{2}\right) m_{s}-6 m_{c}^{2} m_{q}\right) \sqrt{1-\frac{4 m_{c}^{2}}{s}}\, ,
\end{aligned}
\end{equation}

\begin{equation}\nonumber
\rho_{J_{2}}^{6a}(s)=
\frac{\langle\bar{s}s\rangle\langle\bar{q}q\rangle }{16 \pi^{2}}(4m_{c}^{2}+m_{c}m_{q}+m_{c}m_{s})
\sqrt{1-\frac{4 m_{c}^{2}}{s}}\, ,
\end{equation}

\begin{equation}\nonumber
\Pi_{J_{2}}^{6b}\left(M_{B}^{2}\right)=
\frac{\langle\bar{s}s\rangle\langle\bar{q}q\rangle m_{c}^{3}}{16 \pi^{2}} \int_{0}^{1} d \alpha\Big(\frac{m_{q}}{1-\alpha}+\frac{m_{s}}{\alpha}\Big) e^{\frac{-m_{c}^{2}} { \alpha(1-\alpha) M_{B}^{2}}}\, ,
\end{equation}

\begin{equation}\nonumber
\Pi_{J_{2}}^{8}\left(M_{B}^{2}\right)=
\frac{ m_{c}^{4}}{16 \pi^{2}} \int_{0}^{1} d \alpha \frac{ \langle\bar{s}s\rangle \langle\bar{q}g_{s}\sigma\cdot Gq\rangle+\langle\bar{q}q\rangle \langle\bar{s}g_{s}\sigma\cdot Gs\rangle}{(1-\alpha)^{2}M_{B}^{2}} e^{\frac{-m_{c}^{2}} { \alpha(1-\alpha) M_{B}^{2}}}\, ,
\end{equation}

3. Spectral densities for $J_{1\mu}$:\\
\begin{equation}\nonumber
\rho_{J_{1\mu}}^{0a}(s)=
\frac{3} {4096 \pi^{6}}\int_{\alpha_{min}}^{\alpha_{max}} d \alpha \int_{\beta_{min}}^{\beta_{max}} d \beta
 \frac{(1-\alpha-\beta)^{2}} { \alpha^{3} \beta^{3}}(m_{c}^{2}(\alpha+\beta)-\alpha \beta s)^{3}(m_{c}^{2}(\alpha+\beta)-5 \alpha \beta s)\, ,
\end{equation}

\begin{equation}\nonumber
\begin{aligned}
\rho_{J_{1\mu}}^{0b}(s)=&
- \frac{3m_{c}} {1024 \pi^{6}}\int_{\alpha_{min}}^{\alpha_{max}} d \alpha \int_{\beta_{min}}^{\beta_{max}} d \beta
(1-\alpha-\beta)^{2}(m_{c}^{2}(\alpha+\beta)-\alpha \beta s)^{2} \Big(\frac{(2m_{c}^{2}(\alpha+\beta)-5 \alpha \beta s)m_{s}}{\alpha^{3} \beta^{2}}\\
&+\frac{(m_{c}^{2}(\alpha+\beta)-4\alpha \beta s)m_{q}}{\alpha^{2} \beta^{3}}\Big)\, ,
\end{aligned}
\end{equation}

 \begin{equation}\nonumber
 \begin{aligned}
\rho_{J_{1\mu}}^{3a}(s)=&
- \frac{3\langle\bar{s}s\rangle}{256\pi^{4} }
\int_{\alpha_{min}}^{\alpha_{max}} d \alpha \int_{\beta_{min}}^{\beta_{max}} d \beta
(m_{c}^{2}(\alpha+\beta)-\alpha \beta s)\Big[\frac{4(1-\alpha-\beta)(m_{c}^{2}(\alpha+\beta)-2\alpha \beta s)m_{c}} { \alpha^{2} \beta}\\
 &-\frac{4m_{c}^{2}m_{q}+(m_{c}^{2}(\alpha+\beta)-3\alpha \beta s)m_{s}}{\alpha\beta}\Big]\, ,
\end{aligned}
\end{equation}

 \begin{equation}\nonumber
 \begin{aligned}
\rho_{J_{1\mu}}^{3b}(s)=&
- \frac{3\langle\bar{q}q\rangle}{256\pi^{4} }
\int_{\alpha_{min}}^{\alpha_{max}} d \alpha \int_{\beta_{min}}^{\beta_{max}} d \beta
(m_{c}^{2}(\alpha+\beta)-\alpha \beta s)\Big[\frac{2(1-\alpha-\beta)(m_{c}^{2}(\alpha+\beta)-3\alpha \beta s)m_{c}} { \alpha \beta^{2}}\\
 &-\frac{4m_{c}^{2}m_{s}+(m_{c}^{2}(\alpha+\beta)-3\alpha \beta s)m_{q}}{\alpha\beta}\Big]\, ,
\end{aligned}
\end{equation}

\begin{equation}\nonumber
\begin{aligned}
\rho_{J_{1\mu}}^{4a}(s)=&
\frac{\langle g_{s}^{2} G G\rangle m_{c}^{2} }{4096 \pi^{6}}\int_{\alpha_{min}}^{\alpha_{max}} d \alpha \int_{\beta_{min}}^{\beta_{max}} d \beta (1-\alpha-\beta)^{2}
(m_{c}^{2}(\alpha+ \beta)-2 \alpha \beta s)\Big(\frac{1}{\alpha^{3}}+\frac{1}{\beta^{3}}\Big)\, ,
\end{aligned}
\end{equation}

\begin{equation}\nonumber
\begin{aligned}
\rho_{J_{1\mu}}^{4b}(s)=&
\frac{\langle g_{s}^{2} G G\rangle m_{c}^{2} }{4096 \pi^{6}}\int_{\alpha_{min}}^{\alpha_{max}} d \alpha \int_{\beta_{min}}^{\beta_{max}} d \beta (1-\alpha-\beta)
(m_{c}^{2}(\alpha+ \beta)-\alpha \beta s)\Big(\frac{3(m_{c}^{2}(\alpha+ \beta)-3\alpha \beta s)}{\alpha\beta^{2}}\\
&-\frac{(3m_{c}^{2}(\alpha+ \beta)-5\alpha \beta s)}{\alpha^{2}\beta}\Big)\, ,
\end{aligned}
\end{equation}

\begin{equation}\nonumber
\begin{aligned}
\rho_{J_{1\mu}}^{5a}(s)=
&\frac{3 \langle\bar{s} g_{s}\sigma\cdot G s\rangle m_{c}}{256 \pi^{4}} \int_{\alpha_{min}}^{\alpha_{max}} d \alpha \int_{\beta_{min}}^{\beta_{max}} d \beta
 \frac{2m_{c}^{2}(\alpha+\beta)-3  \alpha \beta s} {\alpha}\, ,
\end{aligned}
\end{equation}

\begin{equation}\nonumber
\begin{aligned}
\rho_{J_{1\mu}}^{5b}(s)=
&\frac{3 \langle\bar{q} g_{s}\sigma\cdot G q\rangle m_{c}}{256 \pi^{4}} \int_{\alpha_{min}}^{\alpha_{max}} d \alpha \int_{\beta_{min}}^{\beta_{max}} d \beta
 \Big[( m_{c}^{2}(\alpha+\beta)-2 \alpha \beta s ) \Big(\frac{1}{\beta}-\frac{2(1-\alpha-\beta)}{\beta^{2}}\Big)+\frac{2m_{c}m_{s}}{\beta}\Big]\, ,
\end{aligned}
\end{equation}

\begin{equation}\nonumber
\begin{aligned}
\rho_{J_{1\mu}}^{5c}(s)=
&\frac{\langle\bar{s} g_{s}\sigma\cdot G s\rangle}{768\pi^{4}}\left(\left(s- m_{c}^{2}\right) m_{s}-9 m_{c}^{2} m_{q}\right) \sqrt{1-\frac{4 m_{c}^{2}}{s}}\\
+&\frac{\langle\bar{q} g_{s}\sigma\cdot G q\rangle}{768 \pi^{4}}\left(\left(s- m_{c}^{2}\right) m_{q}-9 m_{c}^{2} m_{s}\right) \sqrt{1-\frac{4 m_{c}^{2}}{s}}\, ,
\end{aligned}
\end{equation}

\begin{equation}\nonumber
\rho_{J_{1\mu}}^{6a}(s)=
\frac{\langle\bar{s}s\rangle\langle\bar{q}q\rangle }{64 \pi^{2}}(4m_{c}^{2}+2m_{c}m_{q}+m_{c}m_{s})
\sqrt{1-\frac{4 m_{c}^{2}}{s}}\, ,
\end{equation}

\begin{equation}\nonumber
\Pi_{J_{1\mu}}^{6b}\left(M_{B}^{2}\right)=
-\frac{\langle\bar{s}s\rangle\langle\bar{q}q\rangle m_{c}^{3}}{32 \pi^{2}} \int_{0}^{1} d \alpha\Big(\frac{m_{s}}{1-\alpha}+\frac{m_{q}}{\alpha}\Big) e^{\frac{-m_{c}^{2}} { \alpha(1-\alpha) M_{B}^{2}}}\, ,
\end{equation}

\begin{equation}\nonumber
\Pi_{J_{1\mu}}^{8}\left(M_{B}^{2}\right)=
\frac{ m_{c}^{4}}{64 \pi^{2}} \int_{0}^{1} d \alpha\Big(\frac{ \langle\bar{q}q\rangle \langle\bar{s}g_{s}\sigma\cdot Gs\rangle+\langle\bar{s}s\rangle \langle\bar{q}g_{s}\sigma\cdot Gq\rangle}{(1-\alpha)^{2}M_{B}^{2}}
-\frac{2 \langle\bar{s}s\rangle \langle\bar{q}g_{s}\sigma\cdot Gq\rangle}{(1-\alpha)m_{c}^{2}}\Big) e^{\frac{-m_{c}^{2}} { \alpha(1-\alpha) M_{B}^{2}}}\, ,
\end{equation}

4. Spectral densities for $J_{2\mu}$:\\
\begin{equation}\nonumber
\rho_{J_{2\mu}}^{0a}(s)=
\frac{3} {4096 \pi^{6}}\int_{\alpha_{min}}^{\alpha_{max}} d \alpha \int_{\beta_{min}}^{\beta_{max}} d \beta
 \frac{(1-\alpha-\beta)^{2}} { \alpha^{3} \beta^{3}}(m_{c}^{2}(\alpha+\beta)-\alpha \beta s)^{3}(m_{c}^{2}(\alpha+\beta)-5 \alpha \beta s)\, ,
\end{equation}

\begin{equation}\nonumber
\begin{aligned}
\rho_{J_{2\mu}}^{0b}(s)=&
- \frac{3m_{c}} {1024 \pi^{6}}\int_{\alpha_{min}}^{\alpha_{max}} d \alpha \int_{\beta_{min}}^{\beta_{max}} d \beta
(1-\alpha-\beta)^{2}(m_{c}^{2}(\alpha+\beta)-\alpha \beta s)^{2} \Big(\frac{(2m_{c}^{2}(\alpha+\beta)-5 \alpha \beta s)m_{q}}{\alpha^{2} \beta^{3}}\\
&+\frac{(m_{c}^{2}(\alpha+\beta)-4\alpha \beta s)m_{s}}{\alpha^{3} \beta^{2}}\Big)\, ,
\end{aligned}
\end{equation}

 \begin{equation}\nonumber
 \begin{aligned}
\rho_{J_{2\mu}}^{3a}(s)=&
- \frac{3\langle\bar{q}q\rangle}{256\pi^{4} }
\int_{\alpha_{min}}^{\alpha_{max}} d \alpha \int_{\beta_{min}}^{\beta_{max}} d \beta
(m_{c}^{2}(\alpha+\beta)-\alpha \beta s)\Big[\frac{4(1-\alpha-\beta)(m_{c}^{2}(\alpha+\beta)-2\alpha \beta s)m_{c}} { \alpha\beta^{2} }\\
 &-\frac{4m_{c}^{2}m_{s}+(m_{c}^{2}(\alpha+\beta)-3\alpha \beta s)m_{q}}{\alpha\beta}\Big]\, ,
\end{aligned}
\end{equation}

 \begin{equation}\nonumber
 \begin{aligned}
\rho_{J_{2\mu}}^{3b}(s)=&
- \frac{3\langle\bar{s}s\rangle}{256\pi^{4} }
\int_{\alpha_{min}}^{\alpha_{max}} d \alpha \int_{\beta_{min}}^{\beta_{max}} d \beta
(m_{c}^{2}(\alpha+\beta)-\alpha \beta s)\Big[\frac{2(1-\alpha-\beta)(m_{c}^{2}(\alpha+\beta)-3\alpha \beta s)m_{c}} { \alpha ^{2} \beta}\\
 &-\frac{4m_{c}^{2}m_{q}+(m_{c}^{2}(\alpha+\beta)-3\alpha \beta s)m_{s}}{\alpha\beta}\Big]\, ,
\end{aligned}
\end{equation}

\begin{equation}\nonumber
\begin{aligned}
\rho_{J_{2\mu}}^{4a}(s)=&
\frac{\langle g_{s}^{2} G G\rangle m_{c}^{2} }{4096 \pi^{6}}\int_{\alpha_{min}}^{\alpha_{max}} d \alpha \int_{\beta_{min}}^{\beta_{max}} d \beta (1-\alpha-\beta)^{2}
(m_{c}^{2}(\alpha+ \beta)-2 \alpha \beta s)\Big(\frac{1}{\alpha^{3}}+\frac{1}{\beta^{3}}\Big)\, ,
\end{aligned}
\end{equation}

\begin{equation}\nonumber
\begin{aligned}
\rho_{J_{2\mu}}^{4b}(s)=&
\frac{\langle g_{s}^{2} G G\rangle m_{c}^{2} }{4096 \pi^{6}}\int_{\alpha_{min}}^{\alpha_{max}} d \alpha \int_{\beta_{min}}^{\beta_{max}} d \beta (1-\alpha-\beta)
(m_{c}^{2}(\alpha+ \beta)-\alpha \beta s)\Big(\frac{3(m_{c}^{2}(\alpha+ \beta)-3\alpha \beta s)}{\alpha^{2} \beta}\\
&-\frac{(3m_{c}^{2}(\alpha+ \beta)-5\alpha \beta s)}{\alpha\beta^{2} }\Big)\, ,
\end{aligned}
\end{equation}

\begin{equation}\nonumber
\begin{aligned}
\rho_{J_{2\mu}}^{5a}(s)=
&\frac{3 \langle\bar{q} g_{s}\sigma\cdot G q\rangle m_{c}}{256 \pi^{4}} \int_{\alpha_{min}}^{\alpha_{max}} d \alpha \int_{\beta_{min}}^{\beta_{max}} d \beta
 \frac{2m_{c}^{2}(\alpha+\beta)-3  \alpha \beta s} {\beta}\, ,
\end{aligned}
\end{equation}

\begin{equation}\nonumber
\begin{aligned}
\rho_{J_{2\mu}}^{5b}(s)=
&\frac{3 \langle\bar{s} g_{s}\sigma\cdot G s\rangle m_{c}}{256 \pi^{4}} \int_{\alpha_{min}}^{\alpha_{max}} d \alpha \int_{\beta_{min}}^{\beta_{max}} d \beta
 \Big[( m_{c}^{2}(\alpha+\beta)-2 \alpha \beta s ) \Big(\frac{1}{\alpha}-\frac{2(1-\alpha-\beta)}{\alpha^{2}}\Big)+\frac{2m_{c}m_{q}}{\alpha}\Big]\, ,
\end{aligned}
\end{equation}

\begin{equation}\nonumber
\begin{aligned}
\rho_{J_{2\mu}}^{5c}(s)=
&\frac{\langle\bar{s} g_{s}\sigma\cdot G s\rangle}{768\pi^{4}}\left(\left(s- m_{c}^{2}\right) m_{s}-9 m_{c}^{2} m_{q}\right) \sqrt{1-\frac{4 m_{c}^{2}}{s}}\\
+&\frac{\langle\bar{q} g_{s}\sigma\cdot G q\rangle}{768 \pi^{4}}\left(\left(s- m_{c}^{2}\right) m_{q}-9 m_{c}^{2} m_{s}\right) \sqrt{1-\frac{4 m_{c}^{2}}{s}}\, ,
\end{aligned}
\end{equation}

\begin{equation}\nonumber
\rho_{J_{2\mu}}^{6a}(s)=
\frac{\langle\bar{s}s\rangle\langle\bar{q}q\rangle }{64 \pi^{2}}(4m_{c}^{2}+2m_{c}m_{s}+m_{c}m_{q})
\sqrt{1-\frac{4 m_{c}^{2}}{s}}\, ,
\end{equation}

\begin{equation}\nonumber
\Pi_{J_{2\mu}}^{6b}\left(M_{B}^{2}\right)=
-\frac{\langle\bar{s}s\rangle\langle\bar{q}q\rangle m_{c}^{3}}{32 \pi^{2}} \int_{0}^{1} d \alpha\Big(\frac{m_{s}}{1-\alpha}+\frac{m_{q}}{\alpha}\Big) e^{\frac{-m_{c}^{2}} { \alpha(1-\alpha) M_{B}^{2}}}\, ,
\end{equation}

\begin{equation}\nonumber
\Pi_{J_{2\mu}}^{8}\left(M_{B}^{2}\right)=
\frac{ m_{c}^{4}}{64 \pi^{2}} \int_{0}^{1} d \alpha\Big(\frac{ \langle\bar{q}q\rangle \langle\bar{s}g_{s}\sigma\cdot Gs\rangle+\langle\bar{s}s\rangle \langle\bar{q}g_{s}\sigma\cdot Gq\rangle}{(1-\alpha)^{2}M_{B}^{2}}
-\frac{2 \langle\bar{q}q\rangle \langle\bar{s}g_{s}\sigma\cdot Gs\rangle}{(1-\alpha)m_{c}^{2}}\Big) e^{\frac{-m_{c}^{2}} { \alpha(1-\alpha) M_{B}^{2}}}\, ,
\end{equation}

5. Spectral densities for $J_{3\mu}$:\\
\begin{equation}\nonumber
\rho_{J_{3\mu}}^{0a}(s)=
\frac{9} {4096 \pi^{6}}\int_{\alpha_{min}}^{\alpha_{max}} d \alpha \int_{\beta_{min}}^{\beta_{max}} d \beta
 \frac{(1-\alpha-\beta)^{2}} { \alpha^{3} \beta^{3}}(m_{c}^{2}(\alpha+\beta)-\alpha \beta s)^{3}(m_{c}^{2}(\alpha+\beta)-5 \alpha \beta s)\, ,
\end{equation}

\begin{equation}\nonumber
\begin{aligned}
\rho_{J_{3\mu}}^{0b}(s)=&
- \frac{9m_{c}} {1024 \pi^{6}}\int_{\alpha_{min}}^{\alpha_{max}} d \alpha \int_{\beta_{min}}^{\beta_{max}} d \beta
(1-\alpha-\beta)^{2}(m_{c}^{2}(\alpha+\beta)-\alpha \beta s)^{2} \Big(\frac{(m_{c}^{2}(\alpha+\beta)-2\alpha \beta s)m_{q}}{\alpha^{3} \beta^{2}}-\frac{ \alpha \beta sm_{s}}{\alpha^{2} \beta^{3}}\Big)\, ,
\end{aligned}
\end{equation}

 \begin{equation}\nonumber
 \begin{aligned}
\rho_{J_{3\mu}}^{3a}(s)=&
\frac{3\langle\bar{s}s\rangle}{256\pi^{4} }
\int_{\alpha_{min}}^{\alpha_{max}} d \alpha \int_{\beta_{min}}^{\beta_{max}} d \beta
(m_{c}^{2}(\alpha+\beta)-\alpha \beta s)\Big[\frac{4(1-\alpha-\beta) s m_{c}} { \beta}+\frac{12m_{c}^{2}m_{q}+3(m_{c}^{2}(\alpha+\beta)-3\alpha \beta s)m_{s}}{\alpha\beta}\Big]\, ,
\end{aligned}
\end{equation}

 \begin{equation}\nonumber
 \begin{aligned}
\rho_{J_{3\mu}}^{3b}(s)=&
- \frac{3\langle\bar{q}q\rangle}{256\pi^{4} }
\int_{\alpha_{min}}^{\alpha_{max}} d \alpha \int_{\beta_{min}}^{\beta_{max}} d \beta
(m_{c}^{2}(\alpha+\beta)-\alpha \beta s)\Big[\frac{2(1-\alpha-\beta)(3m_{c}^{2}(\alpha+\beta)-5\alpha \beta s)m_{c}} { \alpha^{2} \beta}\\
 &-\frac{12m_{c}^{2}m_{s}+3(m_{c}^{2}(\alpha+\beta)-3\alpha \beta s)m_{q}}{\alpha\beta}\Big]\, ,
\end{aligned}
\end{equation}

\begin{equation}\nonumber
\begin{aligned}
\rho_{J_{3\mu}}^{4a}(s)=&
\frac{3\langle g_{s}^{2} G G\rangle m_{c}^{2} }{4096 \pi^{6}}\int_{\alpha_{min}}^{\alpha_{max}} d \alpha \int_{\beta_{min}}^{\beta_{max}} d \beta (1-\alpha-\beta)^{2}
(m_{c}^{2}(\alpha+ \beta)-2 \alpha \beta s)\Big(\frac{1}{\alpha^{3}}+\frac{1}{\beta^{3}}\Big)\, ,
\end{aligned}
\end{equation}

\begin{equation}\nonumber
\begin{aligned}
\rho_{J_{3\mu}}^{4b}(s)=&
\frac{\langle g_{s}^{2} G G\rangle  }{4096 \pi^{6}}\int_{\alpha_{min}}^{\alpha_{max}} d \alpha \int_{\beta_{min}}^{\beta_{max}} d \beta (1-\alpha-\beta)
(m_{c}^{2}(\alpha+ \beta)-\alpha \beta s)\Big(\frac{3m_{c}^{2}(\alpha+ \beta)-5\alpha \beta s}{\alpha\beta^{2}}\\
&-\frac{3(m_{c}^{2}(\alpha+ \beta)-3\alpha \beta s)}{\alpha^{2}\beta}\Big)\, ,
\end{aligned}
\end{equation}

\begin{equation}\nonumber
\begin{aligned}
\rho_{J_{3\mu}}^{5a}(s)=
&-\frac{3 \langle\bar{s} g_{s}\sigma\cdot G s\rangle m_{c}}{256 \pi^{4}} \int_{\alpha_{min}}^{\alpha_{max}} d \alpha \int_{\beta_{min}}^{\beta_{max}} d \beta s \alpha \, ,
\end{aligned}
\end{equation}

\begin{equation}\nonumber
\begin{aligned}
\rho_{J_{3\mu}}^{5b}(s)=
&\frac{ \langle\bar{q} g_{s}\sigma\cdot G q\rangle m_{c}}{256 \pi^{4}} \int_{\alpha_{min}}^{\alpha_{max}} d \alpha \int_{\beta_{min}}^{\beta_{max}} d \beta
\Big[(3 m_{c}^{2}(\alpha+\beta)-4 s \alpha \beta) \Big(\frac{3}{\alpha}+\frac{2(1-\alpha-\beta)}{\alpha^{2}}\Big)-\frac{6m_{c}m_{s}}{\alpha}\Big]\, ,
\end{aligned}
\end{equation}

\begin{equation}\nonumber
\begin{aligned}
\rho_{J_{3\mu}}^{5c}(s)=
&\frac{\langle\bar{q} g_{s}\sigma\cdot G q\rangle}{256\pi^{4}}\left(\left(s- m_{c}^{2}\right) m_{q}-9 m_{c}^{2} m_{s}\right) \sqrt{1-\frac{4 m_{c}^{2}}{s}}\\
+&\frac{\langle\bar{s} g_{s}\sigma\cdot G s\rangle}{256 \pi^{4}}\left(\left(s- m_{c}^{2}\right) m_{s}-9 m_{c}^{2} m_{q}\right) \sqrt{1-\frac{4 m_{c}^{2}}{s}}\, ,
\end{aligned}
\end{equation}

\begin{equation}\nonumber
\rho_{J_{3\mu}}^{6a}(s)=
\frac{3\langle\bar{s}s\rangle\langle\bar{q}q\rangle }{64 \pi^{2}}(4m_{c}^{2}+m_{c}m_{s})
\sqrt{1-\frac{4 m_{c}^{2}}{s}}\, ,
\end{equation}

\begin{equation}\nonumber
\Pi_{J_{3\mu}}^{6b}\left(M_{B}^{2}\right)=
\frac{\langle\bar{s}s\rangle\langle\bar{q}q\rangle m_{c}^{3}}{32 \pi^{2}} \int_{0}^{1} d \alpha\Big(\frac{m_{q}}{1-\alpha}+\frac{m_{s}}{\alpha}\Big) e^{\frac{-m_{c}^{2}} { \alpha(1-\alpha) M_{B}^{2}}}\, ,
\end{equation}

\begin{equation}\nonumber
\Pi_{J_{3\mu}}^{8}\left(M_{B}^{2}\right)=
\frac{ m_{c}^{4}}{64 \pi^{2}} \int_{0}^{1} d \alpha\Big(\frac{ 3(\langle\bar{s}s\rangle \langle\bar{q}g_{s}\sigma\cdot Gq\rangle+\langle\bar{q}q\rangle \langle\bar{s}g_{s}\sigma\cdot Gs\rangle)}{(1-\alpha)^{2}M_{B}^{2}}
+\frac{2 \langle\bar{s}s\rangle \langle\bar{q}g_{s}\sigma\cdot Gq\rangle}{(1-\alpha)m_{c}^{2}}\Big) e^{\frac{-m_{c}^{2}} { \alpha(1-\alpha) M_{B}^{2}}}\, ,
\end{equation}

6. Spectral densities for $J_{4\mu}$:\\
\begin{equation}\nonumber
\rho_{J_{4\mu}}^{0a}(s)=
\frac{9} {4096 \pi^{6}}\int_{\alpha_{min}}^{\alpha_{max}} d \alpha \int_{\beta_{min}}^{\beta_{max}} d \beta
 \frac{(1-\alpha-\beta)^{2}} { \alpha^{3} \beta^{3}}(m_{c}^{2}(\alpha+\beta)-\alpha \beta s)^{3}(m_{c}^{2}(\alpha+\beta)-5 \alpha \beta s)\, ,
\end{equation}

\begin{equation}\nonumber
\begin{aligned}
\rho_{J_{4\mu}}^{0b}(s)=&
- \frac{9m_{c}} {1024 \pi^{6}}\int_{\alpha_{min}}^{\alpha_{max}} d \alpha \int_{\beta_{min}}^{\beta_{max}} d \beta
(1-\alpha-\beta)^{2}(m_{c}^{2}(\alpha+\beta)-\alpha \beta s)^{2} \Big(\frac{(m_{c}^{2}(\alpha+\beta)-2\alpha \beta s)m_{s}}{\alpha^{2} \beta^{3}}-\frac{ \alpha \beta sm_{q}}{\alpha^{3} \beta^{2}}\Big)\, ,
\end{aligned}
\end{equation}

 \begin{equation}\nonumber
 \begin{aligned}
\rho_{J_{4\mu}}^{3a}(s)=&
\frac{3\langle\bar{q}q\rangle}{256\pi^{4} }
\int_{\alpha_{min}}^{\alpha_{max}} d \alpha \int_{\beta_{min}}^{\beta_{max}} d \beta
(m_{c}^{2}(\alpha+\beta)-\alpha \beta s)\Big[\frac{4(1-\alpha-\beta) s m_{c}} { \alpha}+\frac{12m_{c}^{2}m_{s}+3(m_{c}^{2}(\alpha+\beta)-3\alpha \beta s)m_{q}}{\alpha\beta}\Big]\, ,
\end{aligned}
\end{equation}

 \begin{equation}\nonumber
 \begin{aligned}
\rho_{J_{4\mu}}^{3b}(s)=&
- \frac{3\langle\bar{s}s\rangle}{256\pi^{4} }
\int_{\alpha_{min}}^{\alpha_{max}} d \alpha \int_{\beta_{min}}^{\beta_{max}} d \beta
(m_{c}^{2}(\alpha+\beta)-\alpha \beta s)\Big[\frac{2(1-\alpha-\beta)(3m_{c}^{2}(\alpha+\beta)-5\alpha \beta s)m_{c}} { \alpha \beta^{2}}\\
 &-\frac{12m_{c}^{2}m_{q}+3(m_{c}^{2}(\alpha+\beta)-3\alpha \beta s)m_{s}}{\alpha\beta}\Big]\, ,
\end{aligned}
\end{equation}

\begin{equation}\nonumber
\begin{aligned}
\rho_{J_{4\mu}}^{4a}(s)=&
\frac{3\langle g_{s}^{2} G G\rangle m_{c}^{2} }{4096 \pi^{6}}\int_{\alpha_{min}}^{\alpha_{max}} d \alpha \int_{\beta_{min}}^{\beta_{max}} d \beta (1-\alpha-\beta)^{2}
(m_{c}^{2}(\alpha+ \beta)-2 \alpha \beta s)\Big(\frac{1}{\alpha^{3}}+\frac{1}{\beta^{3}}\Big)\, ,
\end{aligned}
\end{equation}

\begin{equation}\nonumber
\begin{aligned}
\rho_{J_{4\mu}}^{4b}(s)=&
\frac{\langle g_{s}^{2} G G\rangle  }{4096 \pi^{6}}\int_{\alpha_{min}}^{\alpha_{max}} d \alpha \int_{\beta_{min}}^{\beta_{max}} d \beta (1-\alpha-\beta)
(m_{c}^{2}(\alpha+ \beta)-\alpha \beta s)\Big(\frac{3m_{c}^{2}(\alpha+ \beta)-5\alpha \beta s}{\alpha^{2}\beta}\\
&-\frac{3(m_{c}^{2}(\alpha+ \beta)-3\alpha \beta s)}{\alpha\beta^{2}}\Big)\, ,
\end{aligned}
\end{equation}

\begin{equation}\nonumber
\begin{aligned}
\rho_{J_{4\mu}}^{5a}(s)=
&-\frac{3 \langle\bar{q} g_{s}\sigma\cdot G q\rangle m_{c}}{256 \pi^{4}} \int_{\alpha_{min}}^{\alpha_{max}} d \alpha \int_{\beta_{min}}^{\beta_{max}} d \beta s \beta \, ,
\end{aligned}
\end{equation}

\begin{equation}\nonumber
\begin{aligned}
\rho_{J_{4\mu}}^{5b}(s)=
&\frac{ \langle\bar{s} g_{s}\sigma\cdot G s\rangle m_{c}}{256 \pi^{4}} \int_{\alpha_{min}}^{\alpha_{max}} d \alpha \int_{\beta_{min}}^{\beta_{max}} d \beta
\Big[(3 m_{c}^{2}(\alpha+\beta)-4 s \alpha \beta) \Big(\frac{3}{\beta}+\frac{2(1-\alpha-\beta)}{\beta^{2}}\Big)-\frac{6m_{c}m_{q}}{\beta}\Big]\, ,
\end{aligned}
\end{equation}

\begin{equation}\nonumber
\begin{aligned}
\rho_{J_{4\mu}}^{5c}(s)=
&\frac{\langle\bar{s} g_{s}\sigma\cdot G s\rangle}{256\pi^{4}}\left(\left(s- m_{c}^{2}\right) m_{s}-9 m_{c}^{2} m_{q}\right) \sqrt{1-\frac{4 m_{c}^{2}}{s}}\\
+&\frac{\langle\bar{q} g_{s}\sigma\cdot G q\rangle}{256 \pi^{4}}\left(\left(s- m_{c}^{2}\right) m_{q}-9 m_{c}^{2} m_{s}\right) \sqrt{1-\frac{4 m_{c}^{2}}{s}}\, ,
\end{aligned}
\end{equation}

\begin{equation}\nonumber
\rho_{J_{4\mu}}^{6a}(s)=
\frac{3\langle\bar{s}s\rangle\langle\bar{q}q\rangle }{64 \pi^{2}}(4m_{c}^{2}+m_{c}m_{q})
\sqrt{1-\frac{4 m_{c}^{2}}{s}}\, ,
\end{equation}

\begin{equation}\nonumber
\Pi_{J_{4\mu}}^{6b}\left(M_{B}^{2}\right)=
\frac{\langle\bar{s}s\rangle\langle\bar{q}q\rangle m_{c}^{3}}{32 \pi^{2}} \int_{0}^{1} d \alpha\Big(\frac{m_{s}}{1-\alpha}+\frac{m_{q}}{\alpha}\Big) e^{\frac{-m_{c}^{2}} { \alpha(1-\alpha) M_{B}^{2}}}\, ,
\end{equation}

\begin{equation}\nonumber
\Pi_{J_{4\mu}}^{8}\left(M_{B}^{2}\right)=
\frac{ m_{c}^{4}}{64 \pi^{2}} \int_{0}^{1} d \alpha\Big(\frac{ 3(\langle\bar{s}s\rangle \langle\bar{q}g_{s}\sigma\cdot Gq\rangle+\langle\bar{q}q\rangle \langle\bar{s}g_{s}\sigma\cdot Gs\rangle)}{(1-\alpha)^{2}M_{B}^{2}}
+\frac{2 \langle\bar{q}q\rangle \langle\bar{s}g_{s}\sigma\cdot Gs\rangle}{(1-\alpha)m_{c}^{2}}\Big) e^{\frac{-m_{c}^{2}} { \alpha(1-\alpha) M_{B}^{2}}}\, ,
\end{equation}

7. Spectral densities for $J_{\mu\nu}$:\\
 \begin{equation}\nonumber
 \begin{aligned}
\rho_{J_{\mu\nu}}^{0a}(s)=
&-\frac{5} {1024 \pi^{6}}\int_{\alpha_{min}}^{\alpha_{max}} d \alpha \int_{\beta_{min}}^{\beta_{max}} d \beta
 \frac{(1-\alpha-\beta)^{2}} { \alpha^{3} \beta^{3}}(m_{c}^{2}(\alpha+\beta)-\alpha \beta s)^{3}\Big((\alpha+\beta+2)(m_{c}^{2}(\alpha+\beta)-\alpha \beta s)\\
 &-3(m_{c}^{2}(\alpha+\beta)-3 \alpha \beta s)\Big)\, 
\end{aligned}
\end{equation}

\begin{equation}\nonumber
\rho_{J_{\mu\nu}}^{0b}(s)=
- \frac{15m_{c}} {512 \pi^{6}}\int_{\alpha_{min}}^{\alpha_{max}} d \alpha \int_{\beta_{min}}^{\beta_{max}} d \beta
(1-\alpha-\beta)^{2}(m_{c}^{2}(\alpha+\beta)-\alpha \beta s)^{2}(m_{c}^{2}(\alpha+\beta)-4 \alpha \beta s) \Big(\frac{m_{s}}{\alpha^{2} \beta^{3}}+\frac{m_{q}}{\alpha^{3} \beta^{2}}\Big)\, ,
\end{equation}

 \begin{equation}\nonumber
 \begin{aligned}
\rho_{J_{\mu\nu}}^{3a}(s)=&
- \frac{15\langle\bar{s}s\rangle}{64\pi^{4} }
\int_{\alpha_{min}}^{\alpha_{max}} d \alpha \int_{\beta_{min}}^{\beta_{max}} d \beta
(m_{c}^{2}(\alpha+\beta)-\alpha \beta s)\Big[\frac{(1-\alpha-\beta)(m_{c}^{2}(\alpha+\beta)-3\alpha \beta s)m_{c}} { \alpha \beta^{2}}\\
 &-\frac{2m_{c}^{2}m_{q}-\alpha \beta s m_{s}+(1-\alpha-\beta)(m_{c}^{2}(\alpha+\beta)-s\alpha\beta)m_{s}}{\alpha\beta}\Big]\, ,
\end{aligned}
\end{equation}

 \begin{equation}\nonumber
 \begin{aligned}
\rho_{J_{\mu\nu}}^{3b}(s)=&
- \frac{15\langle\bar{q}q\rangle}{64\pi^{4} }
\int_{\alpha_{min}}^{\alpha_{max}} d \alpha \int_{\beta_{min}}^{\beta_{max}} d \beta
(m_{c}^{2}(\alpha+\beta)-\alpha \beta s)\Big[\frac{(1-\alpha-\beta)(m_{c}^{2}(\alpha+\beta)-3\alpha \beta s)m_{c}} { \alpha \beta^{2}}\\
 &-\frac{2m_{c}^{2}m_{s}-\alpha \beta s m_{q}+(1-\alpha-\beta)(m_{c}^{2}(\alpha+\beta)-s\alpha\beta)m_{q}}{\alpha\beta}\Big]\, ,
\end{aligned}
\end{equation}

\begin{equation}\nonumber
\begin{aligned}
\rho_{J_{\mu\nu}}^{4a}(s)=
&\frac{5 \langle g_{s}^{2} G G\rangle m_{c}^{2} }{1024 \pi^{6}}\int_{\alpha_{min}}^{\alpha_{max}} d \alpha \int_{\beta_{min}}^{\beta_{max}} d \beta (1-\alpha-\beta)^{2}\Big[
(1-\alpha-\beta)(m_{c}^{2}(\alpha+ \beta)- \alpha \beta s)\Big(\frac{1}{3\alpha^{3}}+\frac{1}{3\beta^{3}}\Big)\\
&-\Big(\frac{\beta s}{2\alpha^{2}}+\frac{\alpha s}{2\beta^{2}}\Big)\Big]\, ,
\end{aligned}
\end{equation}

\begin{equation}\nonumber
\begin{aligned}
\rho_{J_{\mu\nu}}^{4b}(s)=
&\frac{5\langle g_{s}^{2} G G\rangle m_{c}^{2} }{2048 \pi^{6}}\int_{\alpha_{min}}^{\alpha_{max}} d \alpha \int_{\beta_{min}}^{\beta_{max}} d \beta (1-\alpha-\beta)(m_{c}^{2}(\alpha+ \beta)-\alpha \beta s)\Big(\frac{1}{\alpha^{2}\beta}+\frac{1}{\alpha\beta^{2}}\Big)\\
&\times\Big((1-\alpha-\beta)(m_{c}^{2}(\alpha+ \beta)- \alpha \beta s)-4(m_{c}^{2}(\alpha+ \beta)-2 \alpha \beta s)\Big)\, ,
\end{aligned}
\end{equation}

\begin{equation}\nonumber
\begin{aligned}
\rho_{J_{\mu\nu}}^{5a}(s)=
&\frac{5 \langle\bar{s} g_{s}\sigma\cdot G s\rangle m_{c}}{128 \pi^{4}} \int_{\alpha_{min}}^{\alpha_{max}} d \alpha \int_{\beta_{min}}^{\beta_{max}} d \beta
\frac{3( m_{c}^{2}(\alpha+\beta)-2 \alpha \beta s)m_{c}+2( 2m_{c}^{2}(\alpha+\beta)-3 \alpha \beta s)m_{s}}{\beta}\, ,\\
 &+\frac{5 \langle\bar{q} g_{s}\sigma\cdot G q\rangle m_{c}}{128 \pi^{4}} \int_{\alpha_{min}}^{\alpha_{max}} d \alpha \int_{\beta_{min}}^{\beta_{max}} d \beta
\frac{3( m_{c}^{2}(\alpha+\beta)-2 \alpha \beta s)m_{c}+2( 2m_{c}^{2}(\alpha+\beta)-3 \alpha \beta s)m_{q}}{\alpha}\, ,
\end{aligned}
\end{equation}

\begin{equation}\nonumber
\begin{aligned}
\rho_{J_{\mu\nu}}^{5b}(s)=
&\frac{5\langle\bar{q} g_{s}\sigma\cdot G q\rangle}{256 \pi^{4}}\left(\left(s-2 m_{c}^{2}\right) m_{q}-30 m_{c}^{2} m_{s}\right) \sqrt{1-\frac{4 m_{c}^{2}}{s}}\\
+&\frac{5\langle\bar{s} g_{s}\sigma\cdot G s\rangle}{256 \pi^{4}}\left(\left(s-2 m_{c}^{2}\right) m_{s}-30 m_{c}^{2} m_{q}\right) \sqrt{1-\frac{4 m_{c}^{2}}{s}}\, ,
\end{aligned}
\end{equation}

\begin{equation}\nonumber
\rho_{J_{\mu\nu}}^{6a}(s)=
\frac{5\langle\bar{s}s\rangle\langle\bar{q}q\rangle }{32 \pi^{2}}(4m_{c}^{2}+m_{c}m_{q}+m_{c}m_{s})
\sqrt{1-\frac{4 m_{c}^{2}}{s}}\, ,
\end{equation}

\begin{equation}\nonumber
\Pi_{J_{\mu\nu}}^{6b}\left(M_{B}^{2}\right)=
\frac{5\langle\bar{s}s\rangle\langle\bar{q}q\rangle m_{c}^{3}}{16 \pi^{2}} \int_{0}^{1} d \alpha\Big(\frac{m_{q}}{1-\alpha}+\frac{m_{s}}{\alpha}\Big) e^{\frac{-m_{c}^{2}} { \alpha(1-\alpha) M_{B}^{2}}}\, ,
\end{equation}

\begin{equation}\nonumber
\Pi_{J_{\mu\nu}}^{8}\left(M_{B}^{2}\right)=
\frac{ 5m_{c}^{4}}{32 \pi^{2}} \int_{0}^{1} d \alpha \frac{ \langle\bar{s}s\rangle \langle\bar{q}g_{s}\sigma\cdot Gq\rangle+\langle\bar{q}q\rangle \langle\bar{s}g_{s}\sigma\cdot Gs\rangle}{(1-\alpha)^{2}M_{B}^{2}} e^{\frac{-m_{c}^{2}} { \alpha(1-\alpha) M_{B}^{2}}}\, ,
\end{equation}

8. Spectral densities for $\eta_{3\mu}$:\\
\begin{equation}\nonumber
\rho_{\eta_{3\mu}}^{0a}(s)=
\frac{3} {1024 \pi^{6}}\int_{\alpha_{min}}^{\alpha_{max}} d \alpha \int_{\beta_{min}}^{\beta_{max}} d \beta
 \frac{(1-\alpha-\beta)^{2}} { \alpha^{3} \beta^{3}}(m_{c}^{2}(\alpha+\beta)-\alpha \beta s)^{3}(m_{c}^{2}(\alpha+\beta)-5 \alpha \beta s)\, ,
\end{equation}

\begin{equation}\nonumber
\begin{aligned}
\rho_{\eta_{3\mu}}^{0b}(s)=&
- \frac{3m_{c}} {256 \pi^{6}}\int_{\alpha_{min}}^{\alpha_{max}} d \alpha \int_{\beta_{min}}^{\beta_{max}} d \beta
(1-\alpha-\beta)^{2}(m_{c}^{2}(\alpha+\beta)-\alpha \beta s)^{2} \Big(\frac{(m_{c}^{2}(\alpha+\beta)-2\alpha \beta s)m_{s}}{\alpha^{3} \beta^{2}}-\frac{ \alpha \beta sm_{q}}{\alpha^{2} \beta^{3}}\Big)\, ,
\end{aligned}
\end{equation}

 \begin{equation}\nonumber
 \begin{aligned}
\rho_{\eta_{3\mu}}^{3a}(s)=&
\frac{\langle\bar{q}q\rangle}{64\pi^{4} }
\int_{\alpha_{min}}^{\alpha_{max}} d \alpha \int_{\beta_{min}}^{\beta_{max}} d \beta
(m_{c}^{2}(\alpha+\beta)-\alpha \beta s)\Big[\frac{4(1-\alpha-\beta) s m_{c}} { \beta}+\frac{12m_{c}^{2}m_{s}+3(m_{c}^{2}(\alpha+\beta)-3\alpha \beta s)m_{q}}{\alpha\beta}\Big]\, ,
\end{aligned}
\end{equation}

 \begin{equation}\nonumber
 \begin{aligned}
\rho_{\eta_{3\mu}}^{3b}(s)=&
- \frac{\langle\bar{s}s\rangle}{64\pi^{4} }
\int_{\alpha_{min}}^{\alpha_{max}} d \alpha \int_{\beta_{min}}^{\beta_{max}} d \beta
(m_{c}^{2}(\alpha+\beta)-\alpha \beta s)\Big[\frac{2(1-\alpha-\beta)(3m_{c}^{2}(\alpha+\beta)-5\alpha \beta s)m_{c}} { \alpha^{2} \beta}\\
 &-\frac{12m_{c}^{2}m_{q}+3(m_{c}^{2}(\alpha+\beta)-3\alpha \beta s)m_{s}}{\alpha\beta}\Big]\, ,
\end{aligned}
\end{equation}

\begin{equation}\nonumber
\begin{aligned}
\rho_{\eta_{3\mu}}^{4a}(s)=&
\frac{\langle g_{s}^{2} G G\rangle m_{c}^{2} }{1024 \pi^{6}}\int_{\alpha_{min}}^{\alpha_{max}} d \alpha \int_{\beta_{min}}^{\beta_{max}} d \beta (1-\alpha-\beta)^{2}
(m_{c}^{2}(\alpha+ \beta)-2 \alpha \beta s)\Big(\frac{1}{\alpha^{3}}+\frac{1}{\beta^{3}}\Big)\, ,
\end{aligned}
\end{equation}

\begin{equation}\nonumber
\begin{aligned}
\rho_{\eta_{3\mu}}^{4b}(s)=&
\frac{\langle g_{s}^{2} G G\rangle  }{1024 \pi^{6}}\int_{\alpha_{min}}^{\alpha_{max}} d \alpha \int_{\beta_{min}}^{\beta_{max}} d \beta 
(m_{c}^{2}(\alpha+ \beta)-\alpha \beta s)s \Big(3+\frac{4(1-\alpha-\beta)}{\beta}-\frac{3(1-\alpha-\beta)^{2}}{4\beta^{2}}\Big)\, ,
\end{aligned}
\end{equation}

\begin{equation}\nonumber
\begin{aligned}
\rho_{\eta_{3\mu}}^{5a}(s)=
&\frac{ \langle\bar{q}g_{s}\sigma\cdot Gq\rangle m_{c}}{192 \pi^{4}} \int_{\alpha_{min}}^{\alpha_{max}} d \alpha \int_{\beta_{min}}^{\beta_{max}} d \beta
(3 m_{c}^{2}(\alpha+\beta)-4 s \alpha \beta) \left(\frac{1-\alpha+2\beta}{\alpha\beta}\right)\, ,
\end{aligned}
\end{equation}

\begin{equation}\nonumber
\begin{aligned}
\rho_{\eta_{3\mu}}^{5b}(s)=
&-\frac{ \langle\bar{s}g_{s}\sigma\cdot Gs\rangle m_{c}}{384 \pi^{4}} \int_{\alpha_{min}}^{\alpha_{max}} d \alpha \int_{\beta_{min}}^{\beta_{max}} d \beta
\left(1+5\alpha-\beta\right)\, ,
\end{aligned}
\end{equation}

\begin{equation}\nonumber
\begin{aligned}
\rho_{\eta_{3\mu}}^{5c}(s)=
&\frac{\langle\bar{s}g_{s}\sigma\cdot Gs\rangle}{256\pi^{4}}\left(\left(s- m_{c}^{2}\right) m_{s}-9 m_{c}^{2} m_{q}\right) \sqrt{1-\frac{4 m_{c}^{2}}{s}}\\
+&\frac{\langle\bar{q} g_{s}\sigma\cdot Gq\rangle}{256 \pi^{4}}\left(\left(s- m_{c}^{2}\right) m_{q}-9 m_{c}^{2} m_{s}\right) \sqrt{1-\frac{4 m_{c}^{2}}{s}}\, ,
\end{aligned}
\end{equation}

\begin{equation}\nonumber
\rho_{\eta_{3\mu}}^{6a}(s)=
\frac{3\langle\bar{q}q\rangle\langle\bar{s}s\rangle }{16 \pi^{2}}(4m_{c}^{2}+m_{c}m_{q})
\sqrt{1-\frac{4 m_{c}^{2}}{s}}\, ,
\end{equation}

\begin{equation}\nonumber
\Pi_{\eta_{3\mu}}^{6b}\left(M_{B}^{2}\right)=
\frac{\langle\bar{q}q\rangle\langle\bar{s}s\rangle m_{c}^{3}}{24 \pi^{2}} \int_{0}^{1} d \alpha\Big(\frac{m_{s}}{1-\alpha}+\frac{m_{q}}{\alpha}\Big) e^{\frac{-m_{c}^{2}} { \alpha(1-\alpha) M_{B}^{2}}}\, ,
\end{equation}

\begin{equation}\nonumber
\Pi_{\eta_{3\mu}}^{8}\left(M_{B}^{2}\right)=
\frac{ m_{c}^{4}}{96 \pi^{2}} \int_{0}^{1} d \alpha\Big(\frac{ 6(\langle\bar{q}q\rangle \langle\bar{s}g_{s}\sigma\cdot Gs\rangle+\langle\bar{s}s\rangle \bar{q} g_{s}\sigma\cdot Gq\rangle)}{(1-\alpha)^{2}M_{B}^{2}}
+\frac{\langle\bar{q}q\rangle \langle\bar{s}g_{s}\sigma\cdot Gs\rangle+2\langle\bar{s}s\rangle \langle\bar{q} g_{s}\sigma\cdot Gq\rangle}{(1-\alpha)m_{c}^{2}}\Big) e^{\frac{-m_{c}^{2}} { \alpha(1-\alpha) M_{B}^{2}}}\, ,
\end{equation}

9. Spectral densities for $\eta_{4\mu}$:\\
\begin{equation}\nonumber
\rho_{J_{4\mu}}^{0a}(s)=
\frac{3} {1024 \pi^{6}}\int_{\alpha_{min}}^{\alpha_{max}} d \alpha \int_{\beta_{min}}^{\beta_{max}} d \beta
 \frac{(1-\alpha-\beta)^{2}} { \alpha^{3} \beta^{3}}(m_{c}^{2}(\alpha+\beta)-\alpha \beta s)^{3}(m_{c}^{2}(\alpha+\beta)-5 \alpha \beta s)\, ,
\end{equation}

\begin{equation}\nonumber
\begin{aligned}
\rho_{\eta_{4\mu}}^{0b}(s)=&
- \frac{3m_{c}} {256 \pi^{6}}\int_{\alpha_{min}}^{\alpha_{max}} d \alpha \int_{\beta_{min}}^{\beta_{max}} d \beta
(1-\alpha-\beta)^{2}(m_{c}^{2}(\alpha+\beta)-\alpha \beta s)^{2} \Big(\frac{(m_{c}^{2}(\alpha+\beta)-2\alpha \beta s)m_{q}}{\alpha^{3} \beta^{2}}-\frac{ \alpha \beta sm_{s}}{\alpha^{2} \beta^{3}}\Big)\, ,
\end{aligned}
\end{equation}

 \begin{equation}\nonumber
 \begin{aligned}
\rho_{\eta_{4\mu}}^{3a}(s)=&
\frac{\langle\bar{s}s\rangle}{64\pi^{4} }
\int_{\alpha_{min}}^{\alpha_{max}} d \alpha \int_{\beta_{min}}^{\beta_{max}} d \beta
(m_{c}^{2}(\alpha+\beta)-\alpha \beta s)\Big[\frac{4(1-\alpha-\beta) s m_{c}} { \beta}+\frac{12m_{c}^{2}m_{q}+3(m_{c}^{2}(\alpha+\beta)-3\alpha \beta s)m_{s}}{\alpha\beta}\Big]\, ,
\end{aligned}
\end{equation}

 \begin{equation}\nonumber
 \begin{aligned}
\rho_{\eta_{4\mu}}^{3b}(s)=&
- \frac{\langle\bar{q}q\rangle}{64\pi^{4} }
\int_{\alpha_{min}}^{\alpha_{max}} d \alpha \int_{\beta_{min}}^{\beta_{max}} d \beta
(m_{c}^{2}(\alpha+\beta)-\alpha \beta s)\Big[\frac{2(1-\alpha-\beta)(3m_{c}^{2}(\alpha+\beta)-5\alpha \beta s)m_{c}} { \alpha^{2} \beta}\\
 &-\frac{12m_{c}^{2}m_{s}+3(m_{c}^{2}(\alpha+\beta)-3\alpha \beta s)m_{q}}{\alpha\beta}\Big]\, ,
\end{aligned}
\end{equation}

\begin{equation}\nonumber
\begin{aligned}
\rho_{\eta_{4\mu}}^{4a}(s)=&
\frac{\langle g_{s}^{2} G G\rangle m_{c}^{2} }{1024 \pi^{6}}\int_{\alpha_{min}}^{\alpha_{max}} d \alpha \int_{\beta_{min}}^{\beta_{max}} d \beta (1-\alpha-\beta)^{2}
(m_{c}^{2}(\alpha+ \beta)-2 \alpha \beta s)\Big(\frac{1}{\alpha^{3}}+\frac{1}{\beta^{3}}\Big)\, ,
\end{aligned}
\end{equation}

\begin{equation}\nonumber
\begin{aligned}
\rho_{\eta_{4\mu}}^{4b}(s)=&
\frac{\langle g_{s}^{2} G G\rangle  }{1024 \pi^{6}}\int_{\alpha_{min}}^{\alpha_{max}} d \alpha \int_{\beta_{min}}^{\beta_{max}} d \beta 
(m_{c}^{2}(\alpha+ \beta)-\alpha \beta s)s \Big(3+\frac{4(1-\alpha-\beta)}{\beta}-\frac{3(1-\alpha-\beta)^{2}}{4\beta^{2}}\Big)\, ,
\end{aligned}
\end{equation}

\begin{equation}\nonumber
\begin{aligned}
\rho_{\eta_{4\mu}}^{5a}(s)=
&\frac{ \langle\bar{s}g_{s}\sigma\cdot Gs\rangle m_{c}}{192 \pi^{4}} \int_{\alpha_{min}}^{\alpha_{max}} d \alpha \int_{\beta_{min}}^{\beta_{max}} d \beta
(3 m_{c}^{2}(\alpha+\beta)-4 s \alpha \beta) \left(\frac{1-\alpha+2\beta}{\alpha\beta}\right)\, ,
\end{aligned}
\end{equation}

\begin{equation}\nonumber
\begin{aligned}
\rho_{\eta_{4\mu}}^{5b}(s)=
&-\frac{ \langle\bar{q}g_{s}\sigma\cdot Gq\rangle m_{c}}{384 \pi^{4}} \int_{\alpha_{min}}^{\alpha_{max}} d \alpha \int_{\beta_{min}}^{\beta_{max}} d \beta
\left(1+5\alpha-\beta\right)\, ,
\end{aligned}
\end{equation}

\begin{equation}\nonumber
\begin{aligned}
\rho_{\eta_{4\mu}}^{5c}(s)=
&\frac{\langle\bar{s}g_{s}\sigma\cdot Gs\rangle}{256\pi^{4}}\left(\left(s- m_{c}^{2}\right) m_{s}-9 m_{c}^{2} m_{q}\right) \sqrt{1-\frac{4 m_{c}^{2}}{s}}\\
+&\frac{\langle\bar{q} g_{s}\sigma\cdot Gq\rangle}{256 \pi^{4}}\left(\left(s- m_{c}^{2}\right) m_{q}-9 m_{c}^{2} m_{s}\right) \sqrt{1-\frac{4 m_{c}^{2}}{s}}\, ,
\end{aligned}
\end{equation}

\begin{equation}\nonumber
\rho_{\eta_{4\mu}}^{6a}(s)=
\frac{3\langle\bar{q}q\rangle\langle\bar{s}s\rangle }{16 \pi^{2}}(4m_{c}^{2}+m_{c}m_{s})
\sqrt{1-\frac{4 m_{c}^{2}}{s}}\, ,
\end{equation}

\begin{equation}\nonumber
\Pi_{\eta_{4\mu}}^{6b}\left(M_{B}^{2}\right)=
\frac{\langle\bar{q}q\rangle\langle\bar{s}s\rangle m_{c}^{3}}{24 \pi^{2}} \int_{0}^{1} d \alpha\Big(\frac{m_{s}}{1-\alpha}+\frac{m_{q}}{\alpha}\Big) e^{\frac{-m_{c}^{2}} { \alpha(1-\alpha) M_{B}^{2}}}\, ,
\end{equation}

\begin{equation}\nonumber
\Pi_{\eta_{4\mu}}^{8}\left(M_{B}^{2}\right)=
\frac{ m_{c}^{4}}{96 \pi^{2}} \int_{0}^{1} d \alpha\Big(\frac{ 6(\langle\bar{q}q\rangle \langle\bar{s}g_{s}\sigma\cdot Gs\rangle+\langle\bar{s}s\rangle \bar{q} g_{s}\sigma\cdot Gq\rangle)}{(1-\alpha)^{2}M_{B}^{2}}
+\frac{2\langle\bar{q}q\rangle \langle\bar{s}g_{s}\sigma\cdot Gs\rangle+\langle\bar{s}s\rangle \langle\bar{q} g_{s}\sigma\cdot Gq\rangle}{(1-\alpha)m_{c}^{2}}\Big) e^{\frac{-m_{c}^{2}} { \alpha(1-\alpha) M_{B}^{2}}}\, ,
\end{equation}

10. Spectral densities for $\eta_{\mu\nu}$:\\
 \begin{equation}\nonumber
 \begin{aligned}
\rho_{\eta_{\mu\nu}}^{0a}(s)=
&-\frac{5} {768 \pi^{6}}\int_{\alpha_{min}}^{\alpha_{max}} d \alpha \int_{\beta_{min}}^{\beta_{max}} d \beta
 \frac{(1-\alpha-\beta)^{2}} { \alpha^{3} \beta^{3}}(m_{c}^{2}(\alpha+\beta)-\alpha \beta s)^{3}\Big((\alpha+\beta+2)(m_{c}^{2}(\alpha+\beta)-\alpha \beta s)\\
 &-3(m_{c}^{2}(\alpha+\beta)-3 \alpha \beta s)\Big)\, 
\end{aligned}
\end{equation}

\begin{equation}\nonumber
\rho_{\eta_{\mu\nu}}^{0b}(s)=
- \frac{15m_{c}} {384 \pi^{6}}\int_{\alpha_{min}}^{\alpha_{max}} d \alpha \int_{\beta_{min}}^{\beta_{max}} d \beta
(1-\alpha-\beta)^{2}(m_{c}^{2}(\alpha+\beta)-\alpha \beta s)^{2}(m_{c}^{2}(\alpha+\beta)-4 \alpha \beta s) \Big(\frac{m_{s}}{\alpha^{2} \beta^{3}}+\frac{m_{q}}{\alpha^{3} \beta^{2}}\Big)\, ,
\end{equation}

 \begin{equation}\nonumber
 \begin{aligned}
\rho_{\eta_{\mu\nu}}^{3a}(s)=&
- \frac{5\langle\bar{s}s\rangle}{16\pi^{4} }
\int_{\alpha_{min}}^{\alpha_{max}} d \alpha \int_{\beta_{min}}^{\beta_{max}} d \beta
(m_{c}^{2}(\alpha+\beta)-\alpha \beta s)\Big[\frac{(1-\alpha-\beta)(m_{c}^{2}(\alpha+\beta)-3\alpha \beta s)m_{c}} { \alpha^{2} \beta}\\
 &-\frac{2m_{c}^{2}m_{q}-\alpha \beta s m_{s}+(1-\alpha-\beta)(m_{c}^{2}(\alpha+\beta)-s\alpha\beta)m_{s}}{\alpha\beta}\Big]\, ,
\end{aligned}
\end{equation}

 \begin{equation}\nonumber
 \begin{aligned}
\rho_{\eta_{\mu\nu}}^{3b}(s)=&
- \frac{5\langle\bar{q}q\rangle}{16\pi^{4} }
\int_{\alpha_{min}}^{\alpha_{max}} d \alpha \int_{\beta_{min}}^{\beta_{max}} d \beta
(m_{c}^{2}(\alpha+\beta)-\alpha \beta s)\Big[\frac{(1-\alpha-\beta)(m_{c}^{2}(\alpha+\beta)-3\alpha \beta s)m_{c}} { \alpha \beta^{2}}\\
 &-\frac{2m_{c}^{2}m_{s}-\alpha \beta s m_{q}+(1-\alpha-\beta)(m_{c}^{2}(\alpha+\beta)-s\alpha\beta)m_{q}}{\alpha\beta}\Big]\, ,
\end{aligned}
\end{equation}

\begin{equation}\nonumber
\begin{aligned}
\rho_{\eta_{\mu\nu}}^{4a}(s)=
&\frac{5 \langle g_{s}^{2} G G\rangle m_{c}^{2} }{768 \pi^{6}}\int_{\alpha_{min}}^{\alpha_{max}} d \alpha \int_{\beta_{min}}^{\beta_{max}} d \beta (1-\alpha-\beta)^{2}\Big[
(1-\alpha-\beta)(m_{c}^{2}(\alpha+ \beta)- \alpha \beta s)\Big(\frac{1}{3\alpha^{3}}+\frac{1}{3\beta^{3}}\Big)\\
&-\Big(\frac{\beta s}{2\alpha^{2}}+\frac{\alpha s}{2\beta^{2}}\Big)\Big]\, ,
\end{aligned}
\end{equation}

\begin{equation}\nonumber
\begin{aligned}
\rho_{\eta_{\mu\nu}}^{4b}(s)=
&\frac{5\langle g_{s}^{2} G G\rangle m_{c}^{2} }{12288 \pi^{6}}\int_{\alpha_{min}}^{\alpha_{max}} d \alpha \int_{\beta_{min}}^{\beta_{max}} d \beta (m_{c}^{2}(\alpha+ \beta)-\alpha \beta s)\Big[(m_{c}^{2}(\alpha+ \beta)-3\alpha \beta s)\left(1+\frac{2(1-\alpha-\beta)^{2}}{\alpha\beta}\right)\\
&+\frac{4(m_{c}^{2}(\alpha+ \beta)-\alpha \beta s)(1-\alpha-\beta)(\alpha+\beta)}{\alpha\beta^{2}}\Big]\, ,
\end{aligned}
\end{equation}

\begin{equation}\nonumber
\begin{aligned}
\rho_{\eta_{\mu\nu}}^{5a}(s)=
&\frac{5 \langle\bar{s} g_{s}\sigma\cdot G s\rangle m_{c}}{96 \pi^{4}} \int_{\alpha_{min}}^{\alpha_{max}} d \alpha \int_{\beta_{min}}^{\beta_{max}} d \beta
\frac{3( m_{c}^{2}(\alpha+\beta)-2 \alpha \beta s)m_{c}+2( 2m_{c}^{2}(\alpha+\beta)-3 \alpha \beta s)m_{s}}{\beta}\, ,\\
 &+\frac{5 \langle\bar{q} g_{s}\sigma\cdot G q\rangle m_{c}}{96 \pi^{4}} \int_{\alpha_{min}}^{\alpha_{max}} d \alpha \int_{\beta_{min}}^{\beta_{max}} d \beta
\frac{3( m_{c}^{2}(\alpha+\beta)-2 \alpha \beta s)m_{c}+2( 2m_{c}^{2}(\alpha+\beta)-3 \alpha \beta s)m_{q}}{\alpha}\, ,
\end{aligned}
\end{equation}

\begin{equation}\nonumber
\begin{aligned}
\rho_{\eta_{\mu\nu}}^{5b}(s)=
&\frac{5\langle\bar{q} g_{s}\sigma\cdot G q\rangle}{192 \pi^{4}}\left(\left(s-2 m_{c}^{2}\right) m_{q}-30 m_{c}^{2} m_{s}\right) \sqrt{1-\frac{4 m_{c}^{2}}{s}}\\
+&\frac{5\langle\bar{s} g_{s}\sigma\cdot G s\rangle}{192 \pi^{4}}\left(\left(s-2 m_{c}^{2}\right) m_{s}-30 m_{c}^{2} m_{q}\right) \sqrt{1-\frac{4 m_{c}^{2}}{s}}\, ,
\end{aligned}
\end{equation}

\begin{equation}\nonumber
\begin{aligned}
\rho_{\eta_{\mu\nu}}^{5c}(s)=
&\frac{5 \left(\langle\bar{s} g_{s}\sigma\cdot G s\rangle+\langle\bar{q} g_{s}\sigma\cdot G q\rangle\right) m_{c}}{384 \pi^{4}} \int_{\alpha_{min}}^{\alpha_{max}} d \alpha \int_{\beta_{min}}^{\beta_{max}} d \beta
\frac{7(m_{c}^{2}(\alpha+\beta)-6 \alpha \beta s)(\alpha+5(1-\alpha+\beta))}{\alpha\beta}\, ,
\end{aligned}
\end{equation}

\begin{equation}\nonumber
\rho_{\eta_{\mu\nu}}^{6a}(s)=
\frac{5\langle\bar{s}s\rangle\langle\bar{q}q\rangle }{24 \pi^{2}}(4m_{c}^{2}+m_{c}m_{q}+m_{c}m_{s})
\sqrt{1-\frac{4 m_{c}^{2}}{s}}\, ,
\end{equation}

\begin{equation}\nonumber
\Pi_{\eta_{\mu\nu}}^{6b}\left(M_{B}^{2}\right)=
\frac{5\langle\bar{s}s\rangle\langle\bar{q}q\rangle m_{c}^{3}}{12 \pi^{2}} \int_{0}^{1} d \alpha\Big(\frac{m_{q}}{1-\alpha}+\frac{m_{s}}{\alpha}\Big) e^{\frac{-m_{c}^{2}} { \alpha(1-\alpha) M_{B}^{2}}}\, ,
\end{equation}

\begin{equation}\nonumber
\Pi_{\eta_{\mu\nu}}^{8}\left(M_{B}^{2}\right)=
\frac{ 5m_{c}^{4}}{24 \pi^{2}} \int_{0}^{1} d \alpha \left[\frac{ \langle\bar{s}s\rangle \langle\bar{q}g_{s}\sigma\cdot Gq\rangle+\langle\bar{q}q\rangle \langle\bar{s}g_{s}\sigma\cdot Gs\rangle}{(1-\alpha)^{2}M_{B}^{2}}-\frac{\langle\bar{s}s\rangle \langle\bar{q}g_{s}\sigma\cdot Gq\rangle+\langle\bar{q}q\rangle \langle\bar{s}g_{s}\sigma\cdot Gs\rangle}{12\alpha}\right] e^{\frac{-m_{c}^{2}} { \alpha(1-\alpha) M_{B}^{2}}}\, ,
\end{equation}

\end{document}